\documentclass{elsarticle}

\usepackage{lineno,hyperref}
\modulolinenumbers[5]

\usepackage{alltt}

\usepackage{amsmath}
\usepackage{amsfonts}
\usepackage{amssymb}
\usepackage{amsthm}
\usepackage{mathtools}
\usepackage{algorithm}
\usepackage{algcompatible}
\usepackage{tikz}
\usepackage{graphicx,psfrag,epsf}
\usepackage{hyperref}
\usepackage{tabu}

\newcommand{\vect}[1]{\boldsymbol{#1}}
\newcommand{\tp}[1]{{#1}^{\mathsf T}}
\renewcommand{\bar}{\overline}
\newcommand{\eps}{\epsilon}
\newcommand{\pa}{\partial}

\renewcommand{\eps}{\varepsilon}
\renewcommand{\epsilon}{\varepsilon}
\renewcommand{\Sigma}{\varSigma}

\newtheorem{prop}{Proposition}[section]

\newtheorem{pf}{Proof}
\newtheorem{rk}{Remark}
\newcommand{\E}{\mathrm E}
\newcommand{\V}{\mathrm{Var}}
\newcommand{\Cov}{\mathrm{Cov}}
\newcommand{\Cor}{\mathrm{Cor}}
\newcommand{\tr}{\mathrm{tr}}
\DeclareMathAlphabet\mathbfcal{OMS}{cmsy}{b}{n}

\ifpdf
    \graphicspath{{./figure/PNG/}{./figure/PDF/}{./figure/}}
\else
    \graphicspath{{./figure/EPS/}{./figure/}}
\fi

\begin{document}

\begin{frontmatter}

\title{Emulation of Higher-Order Tensors in Manifold Monte Carlo Methods for Bayesian Inverse Problems}


\author[Uwarwickstats]{Shiwei Lan\corref{cor}}
\ead{S.Lan@warwick.ac.uk}

\author[Utexas]{Tan Bui-Thanh}
\ead{tanbui@ices.utexas.edu}

\author[Uheriotwatt]{Mike Christie}
\ead{mike.christie@pet.hw.ac.uk}

\author[Uwarwickstats]{Mark Girolami\corref{cor}}
\ead{M.Girolami@warwick.ac.uk}

\cortext[cor]{Corresponding author}

\address[Uwarwickstats]{Department of Statistics, University of Warwick, Coventry CV4 7AL, UK}
\address[Utexas]{Department of Aerospace Engineering and Engineering Mechanics, Institute for Computational Engineering \& Sciences, The University of Texas, Austin, TX 78712, USA}
\address[Uheriotwatt]{Institute of Petroleum Engineering, Heriot-Watt University, Edinburgh EH14 4AS, UK}

\begin{abstract}
The Bayesian approach to Inverse Problems relies predominantly on Markov Chain Monte Carlo methods for posterior inference. The typical nonlinear concentration of posterior measure observed in many such Inverse Problems presents severe challenges to existing simulation based inference methods.
Motivated by these challenges the exploitation of local geometric information in the form of covariant gradients, metric tensors, Levi-Civita connections, and local geodesic flows, have been introduced to more effectively locally 
explore the configuration space of the posterior measure. However, obtaining such geometric quantities usually requires extensive computational effort and despite their effectiveness affect the applicability of these geometrically-based Monte Carlo methods. In this paper we explore one way to address this issue by the construction of an emulator of the model from which all geometric objects can be obtained in a much more computationally feasible manner. The main concept is to approximate the geometric
quantities using a Gaussian Process emulator which is conditioned on a carefully chosen
design set of configuration points, which also determines the quality of the emulator.
To this end we propose the use of statistical experiment design methods to refine a potentially arbitrarily initialized design
online without destroying the convergence of the resulting Markov chain to the desired invariant measure. The practical examples considered in this paper provide a demonstration of the significant improvement possible in terms of computational loading suggesting this is a promising avenue of further development.
\end{abstract}


\end{frontmatter}



\section{Introduction}
In Bayesian Inverse Problems one needs to draw samples from a typically
high dimensional and complicated intractable probability measure \citep{stuart14}. Samples are
needed to estimate integrals for e.g. point estimates or interval
estimates for uncertainty quantification. Random Walk Metropolis  (RWM) is
hampered with its random walk nature, and Hybrid Monte Carlo (HMC)\citep{duane87,neal10,zhang11,girolami11,shahbaba13,hoffman14,lan14a,betancourt14} with its exploitation of local gradients and approximate Hamiltonian flows in an expanded phase space can substantially improve over RWM. 
Riemannian Manifold Hamiltonian Monte Carlo \citep{girolami11} further 
takes advantage of local metric tensors to adapt the transition kernel of the Markov chain to the local structure of the probability
measure, and indeed the proposal mechanism is provided by the local geodesic flows on the manifold of probability measures \citep{girolami11}. This has been demonstrated to allow Markov Chain Monte Carlo (MCMC) to effectively explore the types of challenging posterior measures observed in many Inverse Problems, see e.g.  \citep{bui14} and the example in Figure \ref{fig:comp4} in this paper.

The challenge here is that these geometric objects including
gradients, metrics, connection components are typically expensive to compute,
hindering their application in practice. This is due to the requirement of a single forward solve of the model in evaluating the likelihood, and this increases with the choice of metric tensor and associated connections (second and third order tensors), see  \citep{bui14} for detailed developments which exploit adjoint solver codes.

In this contribution we investigate the feasibility of emulating these expensive geometric quantities using a Gaussian
Process model \cite{kennedy01}. The remainder of the paper has the following structure. Section 2 briefly reviews Hamiltonian Monte Carlo methods,  Sections 3 and 4 detail the Gaussian Process emulation of potential energies,
gradients, second order metric tensors and third order tensor metric connections. Since it is impossible to emulate the expected Fisher metric \citep{girolami11} based on
the Gaussian Process assumption, we propose to emulate the empirical Fisher information in this work.
The accuracy of the GP emulator to approximate these geometric quantities depends on the design set, or configurations, which should be
well spread over the distribution to capture its geometry. However it is not reasonable to assume such a good design set is available initially.
Therefore section 5 introduces \emph{regeneration} \citep{nummelin84,mykland95,gilks98} as a general adaptation framework and
experimental design algorithm \emph{Mutual Information for Computer Experiments (MICE)} \citep{beck14} to refine the design set.
We illustrate the advantage of emulation for geometric Monte Carlo algorithms over their full versions with examples in section 6.
Finally in section 7, we summarise the contribution and discusse some future directions of investigation.

\section{Review of Dynamics and Geometry Inspired Simulation Methods}
\subsection{HMC}
\emph{Hybrid Monte Carlo (HMC)} \citep{duane87,neal10} is a Metropolis style
algorithm that reduces its random walk behaviour by making distant proposals guided by Hamiltonian flows.
These distant proposals are found by numerically simulating Hamiltonian
dynamics, whose state space consists of its \emph{position} vector,
$\vect\theta\in\mathbb R^D$, and its momentum vector, ${\bf p}\in \mathbb R^D$.
In application to statistical models,
$\vect\theta$ consist of the model parameters (and perhaps latent variables), and ${\bf p}$
are auxiliary variables. The objective is to sample from the posterior
distribution $\pi(\vect\theta|\mathcal D)\propto \pi(\vect\theta)
L(\vect\theta|\mathcal D)$, where $\pi(\vect\theta)$ is the prior and
$L(\vect\theta|\mathcal D)$ is the likelihood function.
We define the \emph{potential energy} as
$U(\vect\theta):=-\log\pi(\vect\theta|\mathcal D)$, and the \emph{kinetic
energy}, $K(p)$, similarly as the minus log of the density of ${\bf p}$, which
is usually assumed ${\bf p}\sim\mathcal N({\bf 0},{\bf M})$. Then the total
energy, \emph{Hamiltonian} function is defined as their sum:
\begin{equation}\label{hamiltonian}
H(\vect\theta, {\bf p}) = U(\vect\theta) + K({\bf p}) = -\log\pi(\vect\theta|\mathcal D) + \frac{1}{2}\tp{\bf p}{\bf M}^{-1}{\bf p}
\end{equation}
Therefore the joint density of $\vect\theta$ and ${\bf p}$ is $\pi(\vect\theta,{\bf p})\propto \exp(-H(\vect\theta, {\bf p}))$.
Note, the covariance matrix ${\bf M}$, also referred as the constant \emph{mass
matrix}.

Given the current state $\vect\theta$, we sample the momentum ${\bf
p}\sim\mathcal N({\bf 0},{\bf M})$, and evolve the joint state ${\bf z}:=(\vect\theta,{\bf p})$
according to \emph{Hamilton's equations}:
\begin{align}\label{HD}
\begin{aligned}
\dot{\vect\theta} &=& \frac{\pa H}{\pa\bf p} &=& {\bf M}^{-1}{\bf p}\\
\dot{\bf p} &=& -\frac{\pa H}{\pa\vect\theta} &=& -\nabla_{\vect\theta}U(\vect\theta)
\end{aligned}
\end{align}
The resulting Hamiltonian dynamics are 1) time reversible, and 2) volume
preserving.
In practice, however, it is difficult to solve Hamiltonian's equations
analytically, so numerical methods such as, \emph{leapfrog} (or St\"ormer-Verlet) \citep{leimkuhler04,hairer06}, to approximate these equations by discretizing
time with small step size $\eps$.  In the standard HMC algorithm, $L$, of these leapfrog steps, with some
step size, $\eps$, are used to propose a new state, which is either accepted or rejected according to the Metropolis acceptance probability
\citep[One can refer to][for more details]{neal10}.

\begin{figure}[t]
  \begin{center}
    \includegraphics[width=1\textwidth,height=.4\textwidth]{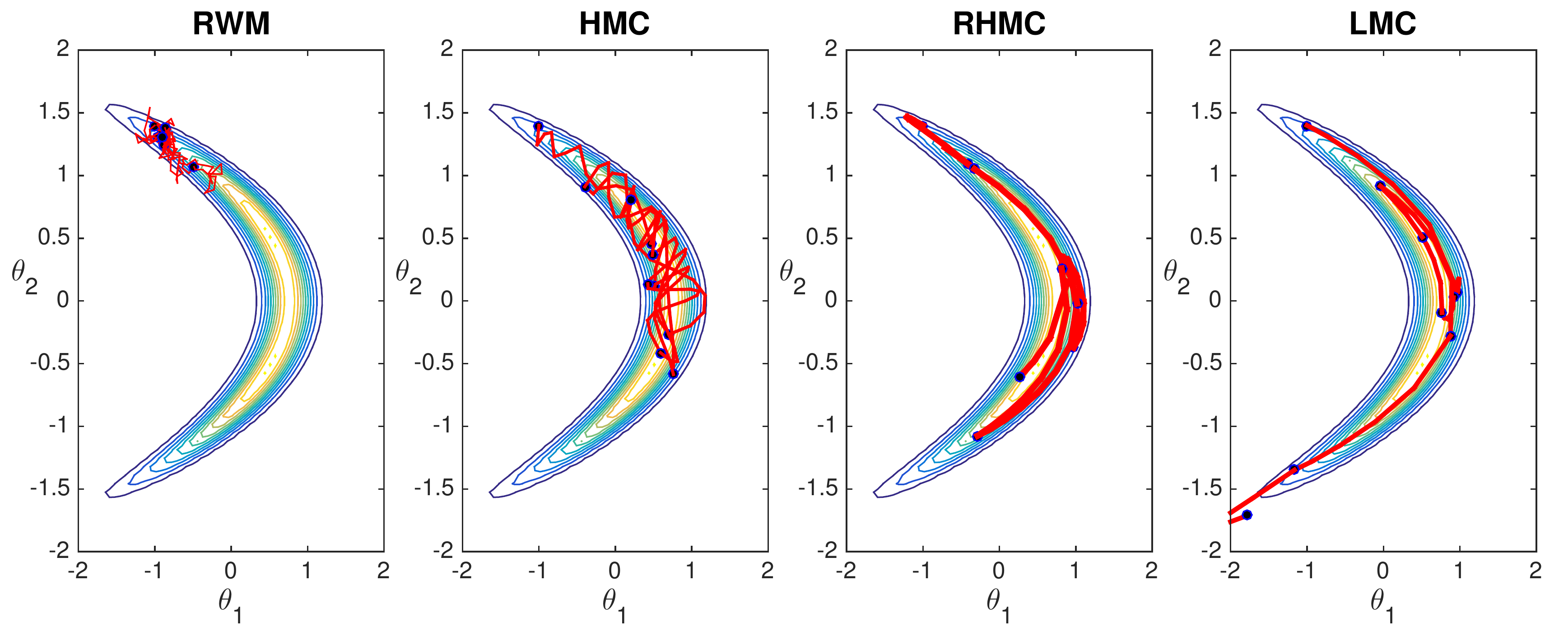}
  \end{center}
  \caption{Comparing Random Walk Metropolis (RWM), Hybrid Monte Carlo (HMC), Riemann Manifold Hamiltonian Monte Carlo (RHMC) and Lagrangian Monte Carlo (LMC) in sampling from a mixture model designed to have a nonlinear concentrated distribution, see \citep{girolami11}. Trajectory length is set to 1.5, the acceptance rates are 0.725,
  0.9, 0.7, 0.8 respectively for the first 10 iterations. Blue dots are
  accepted proposals and red lines are sampling paths, with the thickness
  indicating the computational cost per iteration. It is clear that RWM cannot explore the support of the distribution efficiently while HMC does a much better job, however the global metric structure introduces inefficiencies in exploration, whilst the RHMC and LMC methods with their local geometric structure more effectively traverse the space.}
  \label{fig:comp4}
\end{figure}

\subsection{RHMC}
While HMC explores the parameter space more efficiently than \emph{random walk
Metropolis (RWM)}, it does not fully exploit the geometric properties of the
parameter space. In some complex scenarios, e.g. the concentrated nonlinear
distribution in figure. \ref{fig:comp4}, HMC does not explore the parameter space
as 'straightforwardly' as RHMC does.
To take advantage of the Riemannian geometry of statistical models, \cite{girolami11} proposed
\emph{Riemannian Manifold HMC (RHMC)} to improve the efficiency of the standard HMC by
automatically adapting to the local structure of the parameter space.
Following the argument of \cite{amari00}, they define Hamiltonian dynamics on
the Riemannian manifold endowed with metric tensor ${\bf G}(\vect\theta)$, usually set
to the Fisher information matrix related to the underlying statistical model. As a result, the momentum vector for the
resulting dynamic system becomes ${\bf p}|\vect\theta\sim \mathcal N({\bf 0}, {\bf
G}(\vect\theta))$. That is, the mass matrix ${\bf G}(\vect\theta)$, is
generalized to be position dependent. The Hamiltonian is defined as follows:
\begin{equation}\label{rmhamiltonp}
H({\vect\theta}, {\bf p})  = -\log \pi(\vect\theta|\mathcal D) +\frac12 \log\det{\bf G}(\vect\theta) + 
\frac12 \tp{\bf p}{\bf G}(\vect\theta)^{-1}{\bf p} = \phi(\vect\theta) + \frac12 \tp{\bf p}{\bf G}(\vect\theta)^{-1}{\bf p} 
\end{equation}
where we denote $\phi(\vect\theta):= -\log\pi(\vect\theta|\mathcal D) +\frac12 \log\det{\bf G}(\vect\theta)$.

The resulting Riemannian manifold Hamiltonian dynamics becomes non-separable
since it contains products of $\vect\theta$ and ${\bf p}$. As a result, the
standard leapfrog method is neither time reversible nor volume preserving. To
address this issue, \cite{girolami11} use the following \emph{generalized
leapfrog} method:
\begin{align}
{\bf p}^{(\ell+\frac{1}{2})} 
 & = {\bf p}^{(\ell)} - \frac{\eps}{2} \left[\nabla_{\vect\theta}\phi(\vect\theta^{(\ell)})-\frac12
\vect\nu(\vect\theta^{(\ell)},{\bf p}^{(\ell+\frac{1}{2})})\right] \label{gleapfrog:imp1}\\
\vect\theta^{(\ell+1)} 
 & = \vect\theta^{(\ell)} + \frac{\eps}{2} \left[{\bf G}^{-1}(\vect\theta^{(\ell)}) + {\bf G}^{-1}(\vect\theta^{(\ell+1)})\right]{\bf p}^{(\ell+\frac{1}{2})}\label{gleapfrog:imp2}\\
{\bf p}^{(\ell+1)} 
 & = {\bf p}^{(\ell+\frac{1}{2})} - \frac{\eps}{2}\left[\nabla_{\vect\theta}\phi(\vect\theta^{(\ell+1)})-\frac12
\vect\nu(\vect\theta^{(\ell+1)},{\bf p}^{(\ell+\frac{1}{2})})\right] \label{gleapfrog:xp}
\end{align}
Here, the elements of the vector $(\vect\nu(\vect\theta,{\bf p}))_i = -\tp{\bf
p}\pa_i({\bf G}(\vect\theta)^{-1}){\bf p}$. The above series of transformations
are 1) time reversible, and 2) volume preserving \citep{girolami11}.

\subsection{LMC}
The generalized leapfrog method used for RHMC involves two \emph{implicit}
equations \eqref{gleapfrog:imp1}\eqref{gleapfrog:imp2}, which require potentially
time-consuming fixed-point iterations. In \eqref{gleapfrog:imp2}, it requires
repeatedly inverting the mass matrix ${\bf G}(\vect\theta)$, an $\mathcal
O(D^{2.373})$ operation. To alleviate this problem in \cite{lan12} the authors propose an explicit integrator for RHMC by using the following
{\it Lagrangian} dynamics:
\begin{align*}
\dot{\vect\theta} & = {\bf v} \\
\dot{\bf v} & = -\tp{\bf v}\vect\Gamma(\vect\theta){\bf v} - {\bf G}(\vect\theta)^{-1} \nabla_{\vect\theta}\phi(\vect\theta)
\end{align*}
where the \emph{velocity} ${\bf v}:={\bf G}(\vect\theta)^{-1}{\bf p}\sim\mathcal N({\bf 0}, {\bf G}(\vect\theta)^{-1})$,
and $\vect\Gamma(\vect\theta)$ is the Christoffel Symbols of
the second kind whose $(i,j,k)$-th element is
$\Gamma_{ij}^k=\frac{1}{2}g^{km}(\pa_i g_{mj}+\pa_j g_{im}-\pa_m g_{ij})$ with
$g^{km}$ being the $(k,m)$-th element of ${\bf G}(\vect\theta)^{-1}$.

An \emph{explicit} integrator can be obtained that is time
reversible\label{lan12}:
\begin{align}
{\bf v}^{(\ell+\frac{1}{2})} & = \left[{\bf I} + \frac{\eps}{2} \vect\Omega({\vect\theta}^{(\ell)},{\bf v}^{(\ell)}))\right]^{-1}
\left[{\bf v}^{(\ell)} - \frac{\eps}{2}{\bf G}(\vect\theta^{(\ell)})^{-1} \nabla_{\vect\theta}\phi(\vect\theta^{(\ell)})\right]  \label{lmc:xp1}\\
\vect\theta^{(\ell+1)} &= \vect\theta^{(\ell)} + \eps {\bf v}^{(\ell+\frac{1}{2})}\\
{\bf v}^{(\ell+1)} & = \left[{\bf I} + \frac{\eps}{2} \vect\Omega({\vect\theta}^{(\ell+1)},{\bf v}^{(\ell+\frac{1}{2})}))\right]^{-1}
\left[{\bf v}^{(\ell+\frac{1}{2})} - \frac{\eps}{2}{\bf G}(\vect\theta^{(\ell+1)})^{-1} \nabla_{\vect\theta}\phi(\vect\theta^{(\ell+1)})\right] \label{lmc:xp3}
\end{align}
where $\Omega({\vect\theta}^{(\ell)},{\bf v}^{(\ell)}))_{kj}:=(v^{(\ell)})^i\Gamma(\vect\theta^{(\ell)})_{ij}^k$.
However, the integrator is not volume preserving thus the acceptance probability
is adjusted to ensure the detailed balance condition holds (Denote ${\bf z}:=(\vect\theta,{\bf v})$):
\begin{equation}\label{adjacpt}
\tilde\alpha({\bf z}^{(1)},{\bf z}^{(L+1)}) = \min\left\{1,\exp(-E({\bf z}^{(L+1)})+E({\bf z}^{(1)}))\left|\frac{d{\bf z}^{(L+1)}}{d{\bf z}^{(1)}}\right|\right\}
\end{equation}
where 
$\left|\dfrac{d{\bf z}^{(\ell+1)}}{d{\bf z}^{(\ell)}}\right|=\dfrac{\det({\bf
I}-\eps/2 \vect\Omega({\vect\theta}^{(\ell+1)},{\bf v}^{(\ell+1)})) \det({\bf
I}-\eps/2\vect\Omega({\vect\theta}^{(\ell)},{\bf v}^{(\ell+1/2)}))}{\det({\bf
I}+\eps/2\vect\Omega({\vect\theta}^{(\ell+1)},{\bf v}^{(\ell+1/2)})) \det({\bf
I}+\eps/2\vect\Omega({\vect\theta}^{(\ell)},{\bf v}^{(\ell)}))}$ is the Jacobian determinant, and $E({\bf
z})$ is the \emph{energy} for the Lagrangian dynamics defined as follows:
\begin{equation}\label{energyv}
E({\vect\theta}, {\bf p})  = -\log \pi(\vect\theta|\mathcal D) -\frac12 \log\det{\bf G}(\vect\theta) + 
\frac12 \tp{\bf v}{\bf G}(\vect\theta){\bf v}
\end{equation}
The resulting algorithm, \emph{Lagrangian Monte Carlo (LMC)}, has the advantage of working with a fully explicit integrator, however the determinants of the transformations need to be computed and accounted for to correct for the volume compression of the Lagrangian dynamics \citep[See][for more details]{lan12}.  Independent developments in the molecular dynamics literature arrived at a similar compressive generalised hybrid Monte Carlo method {\cite{fang14}.

In Figure \ref{fig:comp4}, we can see the increasing capability of the methods in exploring a complicated parameter space 
as more geometric information is introduced. However the associated
computational cost increases significantly: RWM is
$\mathcal O(1)$, HMC is $\mathcal O(D)$, and RHMC and LMC are $\mathcal
O(D^3)$ (or $\mathcal O(D^{2.373})$ with faster arithmetic algorithms).
As discussed it is computationally intensive to directly apply these geometric Monte Carlo methods to Bayesian Inverse Problems because of the demanding cost of the required geometric quantities. A previous study where first, second and third order adjoint methods were developed in \cite{bui14} illustrates the computational scaling challenges inherent in these methods. The following section considers statistical emulation as a means of getting around this computational bottleneck.

\section{Statistical Emulation}
Statistical emulation was developed as a computational method from the work on
\emph{Design and Analysis of Computer Experiments (DACE)} in the 1980's {\citep{sacks89,currin91}. It was introduced as
a means of statistical approximation in the simulation of complex process models, with the
application mainly to sensitivity and uncertainty analysis \citep{OHagan06}.
Complex models are developed in the sciences and engineering to simulate the behaviour of complex
physical and natural systems and to reflect the scientific understanding of their
mechanisms. Therefore, such mathematical models or computer programs are often
referred as the \emph{simulator}, denoted as ${\bf y}=f({\bf x})$ with inputs ${\bf
x}$ and outputs ${\bf y}$. The standard techniques of sensitivity and uncertainty
analysis described in \cite{saltelli00} demand repeated model runs, each of
which may be expensive to complete, rendering such methods potentially impractical in practice.
Statistical emulators, on the other hand, built to efficiently approximate the
complex simulators, are much faster to compute and find use in for example climate models \citep{challenor12,challenor13}.
In application to the above sampling methods, we can emulate the required
geometric quantities at much lower cost instead of exactly calculating them.
More specifically, we can predict them using a Gaussian Process conditioned on a set of chosen configurations, see \citep{rasmussen06,neal98,kennedy01,OHagan06,oakley04}.

Gaussian Process (GP) priors are used extensively in statistics \citep{neal98,cressie93} and
machine learning \citep{rasmussen06}. GPs have also become a popular choice in
constructing emulators
\citep{santner03,kennedy01,OHagan06,oakley02,oakley04,kennedy06,conti10}.
MCMC methods have been used alongside emulated models
for inference \citep{kennedy01,OHagan06,oakley02,stephenson10}, however, there is little in the
literature for the use of GP emulation to speed up MCMC methods. As far as we
are aware, \cite{rasmussen03} was the first to use GPs for emulating gradients for HMC, likewise
\cite{bui12} built emulators for MCMC based on a non-stationary Gaussian Process, and we develop these ideas much further in this paper.


\section{Gaussian Process Emulators}
Emulation is based on a set of carefully chosen input points, named
\emph{design points}. These points are chosen to represent the simulator,
usually equally spaced, or uniformly positioned. There are methods such as
\emph{maximin}, \emph{Latin hypercube} \citep{mckay79,morris95} in the
experimental design literature to choose such design points.
In the sampling problem considered, we want design points to be better adapted to the posterior measure, that is, they should evenly scatter over the isocontours. 
In this section, we assume a priori such a design set on which a GP emulator is based.
This assumption will be relaxed in the section that follows.

First, we are interested in emulating $U(\vect\theta)$ as a function of $\vect\theta\in\mathbb R^D$ using Gaussian processes.
Following \cite{stephenson10}, we assume the squared exponential correlation:
\begin{equation}\label{GPU}
\begin{aligned}
U(\cdot) &\sim \mathcal{GP}(\mu(\cdot), \mathcal C(\cdot,\cdot))\\
\mu(\cdot) &= {\bf h}(\cdot)\vect\beta,\quad {\bf h}(\vect\theta):=[1,\tp{\vect\theta},\tp{(\vect\theta^2)}]\\
\mathcal C(\cdot,\cdot) &= \sigma^2{\bf C}(\cdot,\cdot),\quad {\bf C}(\vect\theta^i,\vect\theta^j):=\exp\{-\tp{(\vect\theta^i-\vect\theta^j)}\mathrm{diag}(\vect\rho)(\vect\theta^i-\vect\theta^j)\}
\end{aligned}
\end{equation}
Given a set of $n$ design points $\mathfrak{De}:=\{\vect\theta^1,\cdots,\vect\theta^n\}$,
and conditioned on functional outputs ${\bf u}_{\mathfrak D}:=U(\mathfrak{De})$, we can predict
$U(\vect\theta^*)$ at a set of $m$ new points $\mathfrak E:=\{\vect\theta^{*1},\cdots,\vect\theta^{*m}\}$\footnote{In the standard MCMC setting, we only need prediction at the current state, i.e. $m=1$. But we still use symbol $m$ for the convenience of parallelization.},
denoted as ${\bf u}_{\mathfrak E}$:
\begin{equation}\label{GPpred}
\begin{aligned}
{\bf u}_{\mathfrak E}|{\bf u}_{\mathfrak D},\vect\beta,\sigma^2,\vect\rho &\sim \mathcal N(\vect\mu^*, \sigma^2{\bf C}^*)\\
\vect\mu^* &= {\bf H}_{\mathfrak E}\vect\beta+{\bf C}_{\mathfrak{ED}}{\bf C}_{\mathfrak D}^{-1}({\bf u}_{\mathfrak D}-{\bf H}_{\mathfrak D}\vect\beta)\\
{\bf C}^* &= {\bf C}_{\mathfrak E}-{\bf C}_{\mathfrak{ED}}{\bf C}_{\mathfrak D}^{-1}{\bf C}_{\mathfrak{DE}}
\end{aligned}
\end{equation}
where $\vect\beta$ is a $q$ vector ($q=1+2D$),
${\bf H}_{\mathfrak D}:={\bf h}(\mathfrak{De})$
is an $n\times q$ matrix with $i$-th row ${\bf h}(\vect\theta^{i})$ for $i=1,\cdots,n$, ${\bf H}_{\mathfrak E}={\bf h}(\mathfrak E)$ is an $m\times q$ matrix, ${\bf
C}_{\mathfrak D}:={\bf C}(\mathfrak{De},\mathfrak{De})$ is an $n\times n$ matrix with $(i,j)$-element ${\bf C}(\vect\theta^i,\vect\theta^j)$ as above, ${\bf C}_{\mathfrak{ED}}={\bf C}(\mathfrak E,\mathfrak{De})=\tp{\bf
C}_{\mathfrak{DE}}$ is an $m\times n$ matrix, and ${\bf C}_{\mathfrak E}:={\bf C}(\mathfrak
E,\mathfrak E)$ is an $m\times m$ matrix.
Note in general ${\bf C}_{\mathfrak D}\geq 0$, and in practice we add a small nugget $\nu>0$ to the diagonal of ${\bf C}_{\mathfrak D}$ to ensure it is well conditioned \citep{beck14}.

Given a weak prior for both $\vect\beta$ and $\sigma^2$,
$p(\vect\beta,\sigma^2)\propto \sigma^{-2}$, we can integrate out $\vect\beta$
to obtain \citep{oakley99,stephenson10}:
\begin{equation}\label{GPpred_killbeta}
\begin{aligned}
{\bf u}_{\mathfrak E}|{\bf u}_{\mathfrak D},\sigma^2,\vect\rho &\sim \mathcal N(\vect\mu^{**}, \sigma^2{\bf C}^{**})\\
\vect\mu^{**} &= {\bf H}_{\mathfrak E}\widehat{\vect\beta}+{\bf C}_{\mathfrak{ED}}{\bf C}_{\mathfrak D}^{-1}({\bf u}_{\mathfrak D}-{\bf H}_{\mathfrak D}\widehat{\vect\beta})\\
{\bf C}^{**} &= {\bf C}_{\mathfrak E}-{\bf C}_{\mathfrak{ED}}{\bf C}_{\mathfrak D}^{-1}{\bf C}_{\mathfrak{DE}}\\
&\phantom{=} + ({\bf H}_{\mathfrak E}-{\bf C}_{\mathfrak{ED}}{\bf C}_{\mathfrak D}^{-1}{\bf H}_{\mathfrak D}) (\tp{\bf H}_{\mathfrak D}{\bf C}_{\mathfrak D}^{-1}{\bf H}_{\mathfrak D})^{-1} \tp{({\bf H}_{\mathfrak E}-{\bf C}_{\mathfrak{ED}}{\bf C}_{\mathfrak D}^{-1}{\bf H}_{\mathfrak D})}\\
\widehat{\vect\beta} &=(\tp{\bf H}_{\mathfrak D}{\bf C}_{\mathfrak D}^{-1}{\bf H}_{\mathfrak D})^{-1}\tp{\bf H}_{\mathfrak D}{\bf C}_{\mathfrak D}^{-1}{\bf u}_{\mathfrak D}
=:{\bf P}_{\mathfrak D} {\bf u}_{\mathfrak D}
\end{aligned}
\end{equation}
where ${\bf P}_{\mathfrak D}:={\bf B}_{\mathfrak D}^{-1}\tp{\bf H}_{\mathfrak D}{\bf C}_{\mathfrak
D}^{-1}$ and ${\bf B}_{\mathfrak D}:=\tp{\bf H}_{\mathfrak D}{\bf C}_{\mathfrak D}^{-1}{\bf H}_{\mathfrak D}$.
$\vect\mu^{**} $ is also the Best Linear Unbiased Predictor (BLUP) for ${\bf u}_{\mathfrak E}$ conditioned on ${\bf u}_{\mathfrak D}$ \citep{sacks89}.
We can further integrate out $\sigma^2$ to obtain a T-process \citep{stephenson10}:
\begin{equation}\label{GPpred_killsigma}
\begin{aligned}
{\bf u}_{\mathfrak E}|{\bf u}_{\mathfrak D},\vect\rho &\sim \mathcal T_{n-q}(\vect\mu^{**}, \widehat{\sigma}^2{\bf C}^{**})\\
\widehat{\sigma}^2 &= (n-q-2)^{-1} \tp{({\bf u}_{\mathfrak D}-{\bf H}_{\mathfrak D}\widehat{\vect\beta})} {\bf C}_{\mathfrak D}^{-1} ({\bf u}_{\mathfrak D}-{\bf H}_{\mathfrak D}\widehat{\vect\beta})\\
&= (n-q-2)^{-1} \tp{\bf u}_{\mathfrak D} {\bf C}_{\mathfrak D}^{-1} [{\bf I}-{\bf H}_{\mathfrak D}{\bf P}_{\mathfrak D}] {\bf u}_{\mathfrak D}=:(n-q-2)^{-1} \tp{\bf u}_{\mathfrak D} {\bf Q}_{\mathfrak D} {\bf u}_{\mathfrak D}
\end{aligned}
\end{equation}
where ${\bf Q}_{\mathfrak D}:={\bf C}_{\mathfrak D}^{-1} [{\bf I}-{\bf H}_{\mathfrak D}{\bf P}_{\mathfrak D}]$.
We have the following proposition.
\begin{prop}
\begin{equation}
\begin{aligned}
{\bf P}_{\mathfrak D}{\bf H}_{\mathfrak D}&={\bf I}, & {\bf P}_{\mathfrak D}{\bf C}_{\mathfrak D}&={\bf B}_{\mathfrak D}^{-1} \tp{\bf H}_{\mathfrak D},\\
{\bf Q}_{\mathfrak D} {\bf H}_{\mathfrak D}&={\bf 0}, & {\bf Q}_{\mathfrak D} {\bf C}_{\mathfrak D}&={\bf I} - \tp{\bf P}_{\mathfrak D} \tp{\bf H}_{\mathfrak D},\quad {\bf Q}_{\mathfrak D}=\tp{\bf Q}_{\mathfrak D}, \quad {\bf Q}_{\mathfrak D}>0.
\end{aligned}
\end{equation}
\end{prop}
Denote ${\bf L}_{\mathfrak E}:={\bf H}_{\mathfrak E} {\bf P}_{\mathfrak D} +
{\bf C}_{\mathfrak{ED}} {\bf Q}_{\mathfrak D}$, then we can write
\citep{sacks89}
\begin{equation}\label{linmap}
\begin{aligned}
\vect\mu^{**} &= {\bf L}_{\mathfrak E} {\bf u}_{\mathfrak D}\\
{\bf C}^{**} &= {\bf C}_{\mathfrak E} - {\bf C}_{\mathfrak{ED}}{\bf Q}_{\mathfrak D}{\bf C}_{\mathfrak{DE}} + {\bf H}_{\mathfrak E} {\bf B}_{\mathfrak D}^{-1} \tp{\bf H}_{\mathfrak E} - {\bf H}_{\mathfrak E}{\bf P}_{\mathfrak D}{\bf C}_{\mathfrak{DE}} - \tp{({\bf H}_{\mathfrak E}{\bf P}_{\mathfrak D}{\bf C}_{\mathfrak{DE}})}\\
&= {\bf C}_{\mathfrak E} - \begin{bmatrix}{\bf H}_{\mathfrak E}&{\bf C}_{\mathfrak{ED}}\end{bmatrix} \begin{bmatrix}-{\bf B}_{\mathfrak D}^{-1} & {\bf P}_{\mathfrak D}\\ \tp{\bf P}_{\mathfrak D} & {\bf Q}_{\mathfrak D}\end{bmatrix} \begin{bmatrix}\tp{\bf H}_{\mathfrak E}\\{\bf C}_{\mathfrak{DE}}\end{bmatrix}\\
&= {\bf C}_{\mathfrak E} - \begin{bmatrix}{\bf H}_{\mathfrak E}&{\bf C}_{\mathfrak{ED}}\end{bmatrix} \begin{bmatrix}{\bf 0} & \tp{\bf H}_{\mathfrak D}\\ {\bf H}_{\mathfrak D} & {\bf C}_{\mathfrak D}\end{bmatrix}^{-1} \begin{bmatrix}\tp{\bf H}_{\mathfrak E}\\{\bf C}_{\mathfrak{DE}}\end{bmatrix}
\end{aligned}
\end{equation}

In general $\vect\rho$ cannot be integrated out no matter what prior is given,
therefore in practice it is usually fixed at the Maximum Likelihood Estimate (MLE) or estimated using an appropriate Monte Carlo method
\citep[][and more details in appendix \ref{apdx:MLE}]{andrianakis09}. 
Other parametrizations $\vect\rho={\bf r}^{-2}$ ({\bf r} is called correlation length), $\vect\rho=\mathrm{e}^{-\vect\tau}$
are also used and sometimes preferred because there is no positivity restriction as for $\vect\rho$ in the optimization.
In the end, we use $\vect\mu^{**}$ to predict ${\bf u}_{\mathfrak E}$ with
the associated standard errors $\widehat{\sigma}\sqrt{\mathrm{diag}({\bf
C}^{**})}$.

\subsection{Emulation with derivative information on design points}
\begin{figure}[t]
  \begin{center}
    \includegraphics[width=1\textwidth,height=.4\textwidth]{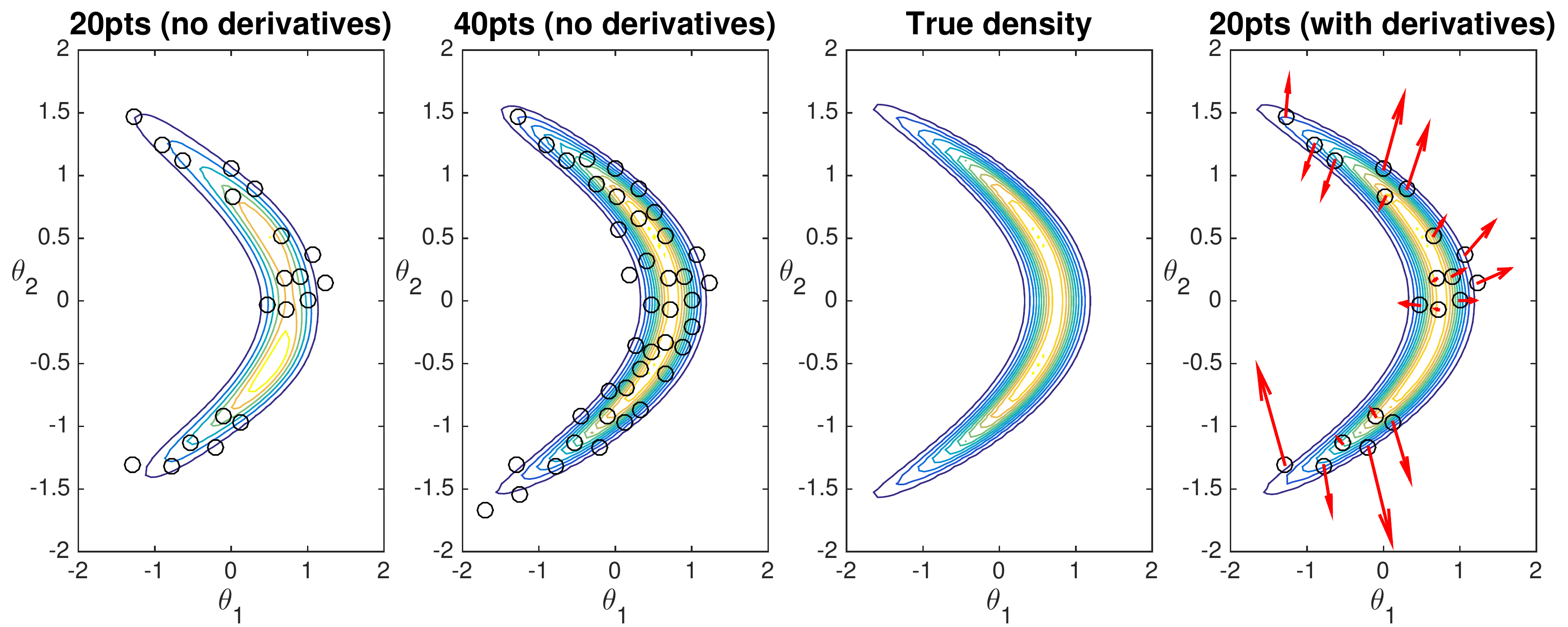}
  \end{center}
  \caption{GP Emulation with and without derivative information on design points. Leftmost:
  density contour estimated without derivative information conditioned on 20 design points ($U(\vect\theta^*)|{\bf u}_{\mathfrak D}, |\mathfrak{De}|=20$);
  Second from left: density contour estimated without derivative information conditioned on 40 design points ($U(\vect\theta^*)|{\bf u}_{\mathfrak D}, |\mathfrak{De}|=40$);
  Second from right: true density contour ($U(\vect\theta^*)$); 
  Rightmost: density contour estimated with derivative information on 20 design points ($U(\vect\theta^*)|\tilde{\bf u}_{\mathfrak D}, |\mathfrak{De}|=20$).}
  \label{fig:gradeff_contour}
\end{figure}

The differential is a linear operator, therefore derivatives of a Gaussian process
are still Gaussian processes given that the mean and covariance functions are differentiable.
This enables us to take advantage of derivative
information on design points, $\nabla U(\mathfrak{De})$, when building the GP emulator, which in
many cases has been shown empirically to improve prediction \citep{stephenson10}.

Following the notations in \cite{stephenson10}, we denote
\begin{equation*}
d{\bf u}_{\mathfrak D}=\nabla\otimes U(\mathfrak{De})=\begin{bmatrix}\frac{\pa U(\mathfrak{De})}{\pa
\theta_1}\\\vdots\\\frac{\pa U(\mathfrak{De})}{\pa \theta_D}\end{bmatrix}_{nD\times 1}, \qquad
\tilde{\bf u}_{\mathfrak D}=\begin{bmatrix}{\bf u}_{\mathfrak D}\\d{\bf u}_{\mathfrak D}\end{bmatrix}_{n(1+D)\times1}
\end{equation*}
where $\otimes$ means Kronecker tensor product. Similarly,
\begin{equation*}
d{\bf H}_{\mathfrak D}=\nabla\otimes {\bf h}(\mathfrak{De}), \qquad \widetilde{\bf H}_{\mathfrak D}=\begin{bmatrix}{\bf
H}_{\mathfrak D}\\d{\bf H}_{\mathfrak D}\end{bmatrix}_{n(1+D)\times q}
\end{equation*}
The differential operator is linear and exchangeable with expectation and
correlation \citep{papoulis02}, therefore
\begin{subequations}\label{diffxint}
\begin{align}
\E\left[\frac{\pa U(\vect\theta^i)}{\pa\theta^i_k}\right] &= \frac{\pa}{\pa\theta^i_k}{\bf h}(\vect\theta^i)\vect\beta\\
& = \beta_k [I(1<k\leq 1+D)+2\theta^i_kI(1+D< k\leq q)] \label{diffxint:dmu}\\
\Cor\left[\frac{\pa U(\vect\theta^i)}{\pa\theta^i_k},U(\vect\theta^j)\right] &= \frac{\pa}{\pa\theta^i_k} {\bf C}(\vect\theta^i,\vect\theta^j) =-2\rho_k(\theta^i_k-\theta^j_k) {\bf C}(\vect\theta^i,\vect\theta^j) \label{diffxint:dC}\\
\Cor\left[\frac{\pa U(\vect\theta^i)}{\pa\theta^i_k},\frac{\pa U(\vect\theta^j)}{\pa\theta^j_l}\right] &= \frac{\pa^2}{\pa\theta^i_k\pa\theta^j_l} {\bf C}(\vect\theta^i,\vect\theta^j)\\
& = [2\rho_k\delta_{kl}-4\rho_k\rho_l(\theta^i_k-\theta^j_k)(\theta^i_l-\theta^j_l)] {\bf C}(\vect\theta^i,\vect\theta^j) \label{diffxint:ddC}
\end{align}
\end{subequations}
Thus we denote ${\bf C}_{\mathfrak D^1\mathfrak D^0}:=\Cor(d{\bf u}_{\mathfrak
D},{\bf u}_{\mathfrak D})_{nD\times n}$, where $\mathfrak D^1$ indicates the first operand
in the correlation operator is the derivative (order 1) of $U$ on $\mathfrak{De}$, 
while $\mathfrak D^0$ means the second operand is $U$ itself (order 0 derivative) on $\mathfrak{De}$.
It has $(ik,j)$-th element as in \eqref{diffxint:dC}.
Similarly ${\bf C}_{\mathfrak D^1\mathfrak D^1}:=\Cor(d{\bf u}_{\mathfrak D},d{\bf u}_{\mathfrak D})_{nD\times nD}$ with
$(ik,jl)$-th element as in \eqref{diffxint:ddC} for $i,j=1,\cdots,n$ and $k,l=1,\cdots,D$.
Finally
\begin{equation*}
\widetilde{\bf C}_{\mathfrak D}=\begin{bmatrix}{\bf C}_{\mathfrak D^0\mathfrak
D^0}&{\bf C}_{\mathfrak D^0\mathfrak D^1}\\{\bf C}_{\mathfrak D^1\mathfrak
D^0}&{\bf C}_{\mathfrak D^1\mathfrak D^1}\end{bmatrix}_{n(1+D)\times n(1+D)}.
\end{equation*}

Now to make prediction based on the extended information that includes the derivatives, i.e. to predict ${\bf
u}_{\mathfrak E}|\tilde{\bf u}_{\mathfrak D}$, we use the same formulae as in
\eqref{linmap} with 
subindices $\mathfrak D$ replaced with $\widetilde{\mathfrak D}$,
that is, ${\bf u}_{\mathfrak D}\leftarrow \tilde{\bf u}_{\mathfrak D}$, ${\bf
H}_{\mathfrak D}\leftarrow \widetilde{\bf H}_{\mathfrak D}$, ${\bf
C}_{\mathfrak D}\leftarrow \widetilde{\bf C}_{\mathfrak D}$ and
${\bf C}_{\mathfrak{ED}}\leftarrow {\bf C}_{\mathfrak E\widetilde{\mathfrak
D}}:=[{\bf C}_{\mathfrak E^0\mathfrak D^0},{\bf C}_{\mathfrak E^0\mathfrak D^1}]_{m\times n(1+D)}
=[\Cor({\bf u}_{\mathfrak E},{\bf u}_{\mathfrak D}),\Cor({\bf u}_{\mathfrak E},d{\bf u}_{\mathfrak D})]$.
Figure \ref{fig:gradeff_contour} shows that with the help of derivative
information on 20 design points, GP emulation (rightmost) is greatly improved in density estimation compared with the one (leftmost) without derivative information, and almost recovers the true density. Even without derivative information, emulation based on 40 design points (second from left) also recovers the truth. This illustrates the tradeoff between using more information on a limited amount of design points and using more design points with a limited amount of additional information.

The effect of gradient information at design points can be explained by the following proposition.
\begin{prop}\label{gradeff}
Conditioned on the same design set $\mathfrak{De}$, Mean Squared Prediction Error (MSPE) of a GP with derivative information is smaller than that of GP without derivative information:
\begin{equation}
\E[(U(\vect\theta^*)-\hat U(\vect\theta^*)|\tilde{\bf u}_{\mathfrak D})^2] \leq \E[(U(\vect\theta^*)-\hat U(\vect\theta^*)|{\bf u}_{\mathfrak D})^2]
\end{equation}
\end{prop}
\begin{pf}
See \ref{apdx:gradeff}.
\end{pf}
Similarly the effect of the number of design points on functional output can be explained by the following  proposition \citep{haaland14}.
\begin{prop}[Benjamin Haaland, Vaibhav Maheshwari]\label{neff}
For design sets $\mathfrak{De}_1\subseteq \mathfrak{De}_2$, we have
\begin{equation}
\E[(U(\vect\theta^*)-\hat U(\vect\theta^*|\mathfrak{De}_2))^2] \leq \E[(U(\vect\theta^*)-\hat U(\vect\theta^*|\mathfrak{De}_1))^2]
\end{equation}
\end{prop}
\begin{rk}
One can think of GP emulation as approximating an infinite dimensional function $U(\vect\theta^*)$ with a vector in finite dimensional space $\mathrm{span}({\bf u}_{\mathfrak D})$ (or $\mathrm{span}(\tilde{\bf u}_{\mathfrak D})$). The approximation error in a finite dimensional vector space $\mathrm{span}(\tilde{\bf u}_{\mathfrak D})$( or $\mathrm{span}({\bf u}_{\mathfrak D_2}))$ is always smaller than that in its subspace $\mathrm{span}({\bf u}_{\mathfrak D})$ (or $\mathrm{span}({\bf u}_{\mathfrak D_1})$).

Heuristically, proposition \ref{gradeff} and proposition \ref{neff} mean that the prediction will be more credible with more information incorporated. Such extra information comes from either derivatives in addition to function values at design points or function values at more design points.
\end{rk}

At the end of this subsection, we comment on the derivative information at design points, $d{\bf u}_{\mathfrak D}$. It is not required for our development of methodology, yet in many scenarios it is not available, e.g. oil reservoir simulation. When this is the case, we relax our notation to let terms with tilde still denote quantities without such derivative information. In the following, we denote
\begin{equation*}
\tilde n=\begin{dcases}n,& d{\bf u}_{\mathfrak D}\; absent\\n(1+D),& d{\bf u}_{\mathfrak D}\; present\end{dcases}.
\end{equation*}

\subsection{Emulating gradients}
Now we want to predict the gradients of the potential at new points for the
use in HMC based algorithms, i.e. to predict $d{\bf u}_{\mathfrak E}|\tilde{\bf u}_{\mathfrak D}$.
The prediction formulae are similar as above
\eqref{linmap} with all the $\mathfrak E^{(0)} $ in subindices replaced with $\mathfrak E^1$, to indicate emulating derivatives on new points $\mathfrak E$.
To be more specific, we write down the conditional mean and correlation:
\begin{equation}\label{GPpredgrad}
\begin{aligned}
\vect\mu^{**} &= {\bf L}_{\mathfrak E^1} \tilde{\bf u}_{\mathfrak D}\\
{\bf C}^{**} &= {\bf C}_{\mathfrak E^1} - {\bf C}_{\mathfrak E^1\mathfrak{\widetilde D}}\widetilde{\bf Q}_{\mathfrak D}{\bf C}_{\mathfrak{\widetilde D}\mathfrak E^1} + {\bf H}_{\mathfrak E^1} \widetilde{\bf B}_{\mathfrak D}^{-1} \tp{\bf H}_{\mathfrak E^1} - {\bf H}_{\mathfrak E^1}\widetilde{\bf P}_{\mathfrak D}{\bf C}_{\mathfrak{\widetilde D}\mathfrak E^1} - \tp{({\bf H}_{\mathfrak E^1}\widetilde{\bf P}_{\mathfrak D}{\bf C}_{\mathfrak{\widetilde D}\mathfrak E^1})}
\end{aligned}
\end{equation}
where ${\bf H}_{\mathfrak E^1}=d{\bf H}_{\mathfrak E}$, ${\bf C}_{\mathfrak E^1\mathfrak{\widetilde D}}:=\Cor(d{\bf u}_{\mathfrak E},\tilde{\bf u}_{\mathfrak D})_{mD\times\tilde n}$.

One can see from the left panel of figure \ref{fig:emu}, in this
illustrative example of a banana shaped distribution, with more and more design points well
positioned, the emulated gradients become closer and closer to the true
gradients and with 30 design points spread over the density contour, the
emulated gradients are close approximations to the true gradients.
We can modify proposition \ref{neff} to show the same effect of design size on emulation of gradients:
\begin{prop}
For design sets $\mathfrak{De}_1\subseteq \mathfrak{De}_2$, we have
\begin{equation*}
\E[(dU(\vect\theta^*)-\widehat{dU}(\vect\theta^*|\mathfrak{De}_2))^2] \leq \E[(dU(\vect\theta^*)-\widehat{dU}(\vect\theta^*|\mathfrak{De}_1))^2]
\end{equation*}
\end{prop}

\begin{figure}[t]
  \begin{center}
     \includegraphics[width=1\textwidth,height=.45\textwidth]{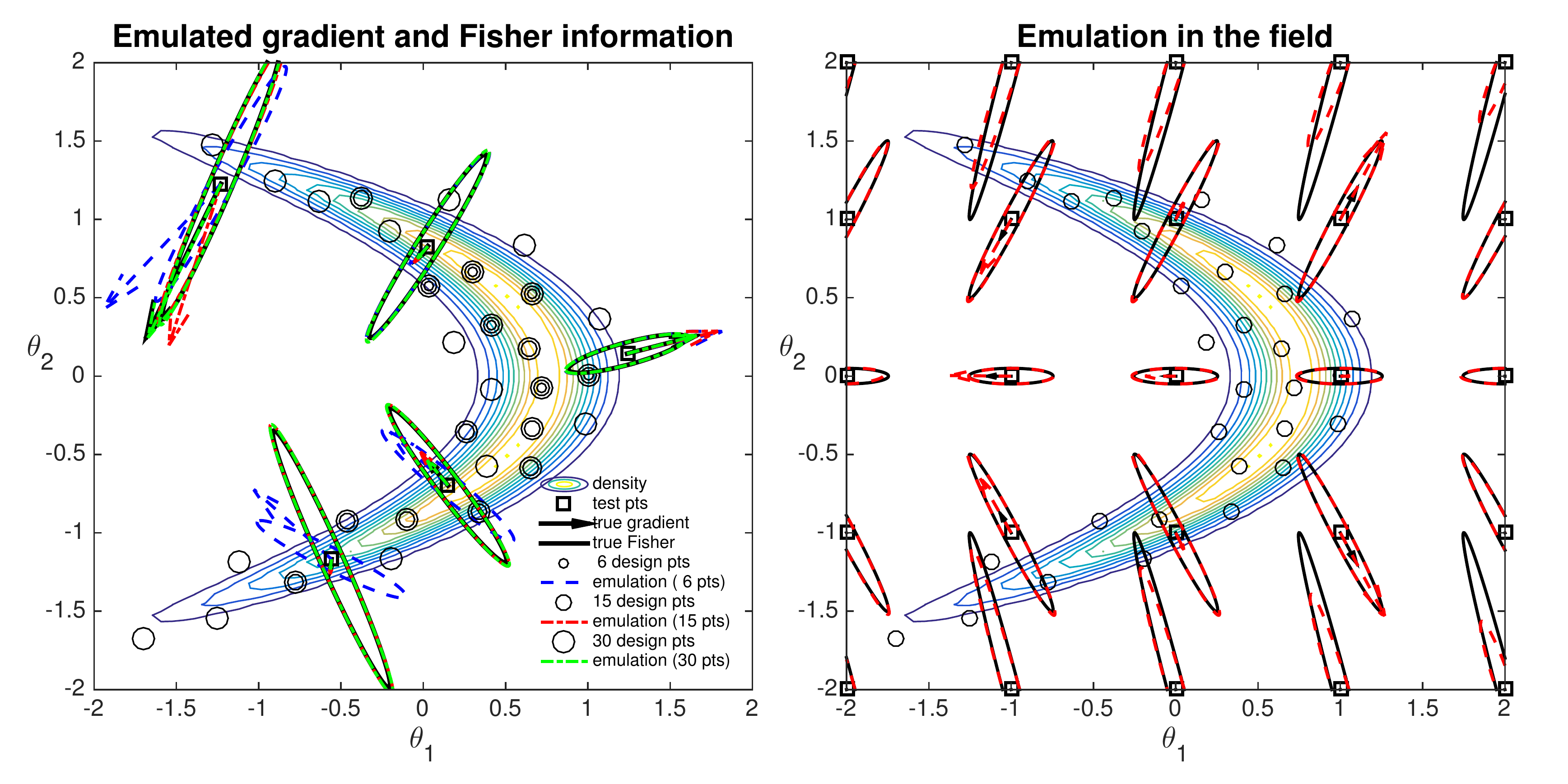}
  \end{center}
  \caption{Emulation using GP. Left: emulated gradients/metrics approximating
  true gradients/metrics; Right: emulations on a $5\times 5$ grid mesh.}
  \label{fig:emu}
\end{figure}

\subsection{Emulating Hessians}
We also need to emulate the Hessian matrix based on the extended
information $\tilde{\bf u}_{\mathfrak D}$.
First, we need to vectorize the Hessian matrix
$[\frac{\pa^2}{\pa\theta_k\pa\theta_l}U(\vect\theta)]$. Denote the vectorized
Hessian matrices evaluated at the new points $\mathfrak E$ as $d^2{\bf
u}_{\mathfrak E}:=\nabla\otimes\nabla\otimes{\bf u}(\mathfrak E)$.
The formulae for prediction remain the same as in \eqref{GPpredgrad} except that
superindices 1 appearing in $\mathfrak E^1$ will be replaced with 2.
${\bf H}_{\mathfrak E^2}=(d^2{\bf H}_{\mathfrak E})_{mD^2\times q}$. 
${\bf C}_{\mathfrak E^2\mathfrak{\widetilde D}}:=\Cor(d^2{\bf u}_{\mathfrak E},\tilde{\bf
u}_{\mathfrak D})_{mD^2\times\tilde n}$. Note
$(ikl,j)$-th element of ${\bf C}_{\mathfrak E^2\mathfrak D^0}$ and $(ikl,jm)$-th element of ${\bf C}_{\mathfrak E^2\mathfrak D^1}$ ($d{\bf u}_{\mathfrak D}$ present) can be deduced from \cite{papoulis02} similarly ($i,j=1,\cdots,n$ and $k,l,m=1,\cdots,D$):
\begin{equation}\label{highercorr}
\begin{aligned}
\Cor\left[\frac{\pa^2 U(\vect\theta^{*i})}{\pa\theta^{*i}_k\pa\theta^{*i}_l},U(\vect\theta^j)\right] &= \frac{\pa^2}{\pa\theta^{*i}_k\pa\theta^{*i}_l} {\bf C}(\vect\theta^{*i},\vect\theta^j)\\
& = [-2\rho_k\delta_{kl}+4\rho_k\rho_l(\theta^{*i}_k-\theta^j_k)(\theta^{*i}_l-\theta^j_l)] {\bf C}(\vect\theta^{*i},\vect\theta^j)\\
\Cor\left[\frac{\pa^2 U(\vect\theta^{*i})}{\pa\theta^{*i}_k\pa\theta^{*i}_l},\frac{\pa U(\vect\theta^j)}{\pa\theta^j_m}\right] &= \frac{\pa^3}{\pa\theta^{*i}_k\pa\theta^{*i}_l\pa\theta^j_m} {\bf C}(\vect\theta^{*i},\vect\theta^j)\\
& = [\boxed{-4\delta_{kl}\rho_m\rho_k(\theta^{*i}_m-\theta^j_m)}-4\delta_{lm}\rho_k\rho_l(\theta^{*i}_k-\theta^j_k)\\
&\phantom{=\,} -4\delta_{mk}\rho_l\rho_m(\theta^{*i}_l-\theta^j_l)\\
&+8\rho_k\rho_l\rho_m(\theta^{*i}_k-\theta^j_k)(\theta^{*i}_l-\theta^j_l)(\theta^{*i}_m-\theta^j_m)] {\bf C}(\vect\theta^{*i},\vect\theta^j)\\
\end{aligned}
\end{equation}

\begin{table}[htbp]
\centering
\begin{tabu}{|[1.2pt]c|[1.2pt]c|c|c|[2pt]c|c|[1.2pt]}
  \tabucline[1.2pt]{-}
& $d^2{\bf u}_{\mathfrak E}\,(mD^2)$ & $d{\bf u}_{\mathfrak E}\,(mD)$ & ${\bf u}_{\mathfrak E}\,(m)$ &
${\bf u}_{\mathfrak D}\,(n)$ & $d{\bf u}_{\mathfrak D}\,(nD)$ \\
  \tabucline[1.2pt]{-}
 $d^2{\bf u}_{\mathfrak E}\,(mD^2)$ & ${\bf C}_{\mathfrak E^2}$ &  &  & ${\bf
 C}_{\mathfrak E^2\mathfrak D^0}$ & ${\bf C}_{\mathfrak E^2\mathfrak
 D^1}$\\
  \hline
 $d{\bf u}_{\mathfrak E}\,(mD)$ &  & ${\bf C}_{\mathfrak E^1}$ &  & ${\bf
 C}_{\mathfrak E^1\mathfrak D^0}$ & ${\bf C}_{\mathfrak E^1\mathfrak D^1}$\\
  \hline
 ${\bf u}_{\mathfrak E}\,(m)$ &  &  & ${\bf C}_{\mathfrak E^0}$ & ${\bf
 C}_{\mathfrak E^0\mathfrak D^0}$ & ${\bf C}_{\mathfrak E^0\mathfrak D^1}$\\
 \tabucline[2pt]{-}
 ${\bf u}_{\mathfrak D}\,(n)$ & ${\bf C}_{\mathfrak D^0\mathfrak E^2}$ & ${\bf
 C}_{\mathfrak D^0\mathfrak E^1}$ & ${\bf C}_{\mathfrak D^0\mathfrak E^0}$ &
 ${\bf C}_{\mathfrak D^0\mathfrak D^0}$ & ${\bf C}_{\mathfrak D^0\mathfrak D^1}$
 \\
 \hline
 $d{\bf u}_{\mathfrak D}\,(nD)$ & ${\bf C}_{\mathfrak D^1\mathfrak E^2}$ & ${\bf
 C}_{\mathfrak D^1\mathfrak E^1}$ & ${\bf C}_{\mathfrak D^1\mathfrak E^0}$ &
 ${\bf C}_{\mathfrak D^1\mathfrak D^0}$ & ${\bf C}_{\mathfrak D^1\mathfrak D^1}$
 \\
  \tabucline[1.2pt]{-}
\end{tabu}
\caption{Relationship between blocks of the whole correlation matrix.} 
\label{tab:corrmatrix}
\end{table}
Table \eqref{tab:corrmatrix} may help better understand
the relationship between blocks of the correlation matrix.
To sum up, we denote $\widetilde{\bf B}_{\mathfrak
D}:=\tp{\widetilde{\bf H}}_{\mathfrak D}\widetilde{\bf C}_{\mathfrak
D}^{-1}\widetilde{\bf H}_{\mathfrak D}$, $\widetilde{\bf P}_{\mathfrak
D}:=\widetilde{\bf B}_{\mathfrak D}^{-1}\tp{\widetilde{\bf H}}_{\mathfrak
D}\widetilde{\bf C}_{\mathfrak D}^{-1}$, and $\widetilde{\bf Q}_{\mathfrak
D}:=\widetilde{\bf C}_{\mathfrak D}^{-1}({\bf I}-\widetilde{\bf H}_{\mathfrak
D}\widetilde{\bf P}_{\mathfrak D})$. Then all the predictions can be
cast as a linear mapping of the extended information at designed
points, $\tilde{\bf u}_{\mathfrak D}$:
\begin{equation}\label{predsumry}
\E[{\bf u}_{\mathfrak E^{\alpha}}|\tilde{\bf u}_{\mathfrak D}]=\widetilde{\bf L}_{\mathfrak E^{\alpha}} \tilde{\bf u}_{\mathfrak D},\quad \widetilde{\bf L}_{\mathfrak E^{\alpha}}:= {\bf H}_{\mathfrak E^{\alpha}}\widetilde{\bf P}_{\mathfrak D}+{\bf C}_{\mathfrak E^{\alpha}\mathfrak{\widetilde D}}\widetilde{\bf Q}_{\mathfrak D},\quad \alpha=0,1,2
\end{equation}

\subsection{Emulating Fisher information}
Last but not the least, the Fisher metric and Christoffel symbols require to be emulated
for RHMC and LMC. The expected Fisher information involves the following expectation with respect to ${\bf x}$, not $\vect\theta$:
\begin{equation}\label{FIint}
\mathrm{FI}(\vect\theta^*|\mathfrak{De}) = \int \nabla^2 U({\bf x},\vect\theta^*|\mathfrak{De}) \exp(-U({\bf x},\vect\theta^*|\mathfrak{De})) d{\bf x}
\end{equation}
whose integrand is no longer a GP under assumption \eqref{GPU}, thus direct
emulation of the expected Fisher information using a GP is not readily available.

We instead consider the \emph{empirical Fisher information}, which can be a good estimate
of expected Fisher information when there are sufficient data:
\begin{equation}\label{eFI}
\begin{split}
\mathrm{eFI}(\vect\theta^*|\mathfrak{De}) &= \mathrm D U(\vect\theta^*|\mathfrak{De}) \tp{\mathrm D U(\vect\theta^*|\mathfrak{De})}  - \frac{1}{N} \nabla U(\vect\theta^*|\mathfrak{De}) \tp{\nabla U(\vect\theta^*|\mathfrak{De})}\\
&= \mathrm D U(\vect\theta^*|\mathfrak{De}) [{\bf I}_N-{\bf 1}_N\tp{\bf 1}_N/N]\tp{\mathrm D U(\vect\theta^*|\mathfrak{De})} =: \mathrm D U(\vect\theta^*|\mathfrak{De}) {\bf J}_N \tp{\mathrm D U(\vect\theta^*|\mathfrak{De})}
\end{split}
\end{equation}
where $N$ is the number of data, $\mathrm D U(\vect\theta^*|\mathfrak{De})$ is a $D\times N$ matrix whose
$(i,j)$-th element is $\frac{\pa}{\pa\theta^*_i} U({\bf x}_j,\vect\theta^*|\mathfrak{De})$, and 
$\nabla U(\vect\theta^*|\mathfrak{De})=\mathrm D U(\vect\theta^*|\mathfrak{De})
{\bf 1}_N$ is a $D$ vector whose $i$-th element is $\frac{\pa}{\pa\theta^*_i}
U({\bf X},\vect\theta^*|\mathfrak{De})=\sum_{j=1}^N\frac{\pa}{\pa\theta^*_i}U({\bf x}_j,\vect\theta^*|\mathfrak{De})$.
${\bf J}_N:={\bf I}_N-{\bf 1}_N\tp{\bf 1}_N/N$.
Now we focus on how to estimate $\mathrm D U(\vect\theta^*|\mathfrak{De})$,
instead of $\nabla U(\vect\theta^*|\mathfrak{De})$.

It is still impossible to estimate $\frac{\pa}{\pa\theta^*_i} U({\bf
x}_j,\vect\theta^*|\mathfrak{De})$ with assumption \eqref{GPU}, so we relax it
to assume the same GP for $U({\bf x}_j,\cdot)$ across different ${\bf x}_j$'s:
\begin{equation}\label{GPUi}
U({\bf x}_j,\,\cdot) \sim \mathcal{GP}(\mu(\cdot), \mathcal C(\cdot,\cdot))\\
\end{equation}
Denote ${\bf U}_{\mathfrak D}:=U({\bf X},\mathfrak{De})$ as an $n\times N$
matrix with $(i,j)$-th element $U({\bf x}_j,\vect\theta^i)$ and
$d{\bf U}_{\mathfrak D}=\nabla\otimes U({\bf X},\mathfrak{De})$ as an $nD\times
N$ matrix if available. Let $\widetilde{\bf U}_{\mathfrak D}=\tp{[\tp{\bf U}_{\mathfrak D},\tp{d{\bf U}}_{\mathfrak D}]}$ denote an $\tilde n\times N$ matrix.
Similarly $d{\bf U}_{\mathfrak E}=\nabla\otimes U({\bf X},\mathfrak E)$ denotes
an $mD\times N$ matrix.

Applying a similar argument to each column of $\widetilde{\bf U}_{\mathfrak D}$, we obtain the linear prediction \eqref{predsumry} for
$d{\bf U}_{\mathfrak E}|\widetilde{\bf U}_{\mathfrak D}$:
\begin{equation}\label{GPpredGRAD}
\E[d{\bf U}_{\mathfrak E}|\widetilde{\bf U}_{\mathfrak D}] = \widetilde{\bf L}_{\mathfrak E^1}\widetilde{\bf U}_{\mathfrak D}
\end{equation}
To predict the empirical Fisher information on the evaluation set $\mathfrak E$,
$\mathrm{eFI}_{\mathfrak E}|\widetilde{\bf U}_{\mathfrak D}$, we
substitute $\mathrm D U({\mathfrak E}|\mathfrak{De})\approx \widetilde{\bf L}_{\mathfrak E^1}\widetilde{\bf U}_{\mathfrak D}$ back in \eqref{eFI} to get the following $mD\times mD$
matrix\footnote{To get $\mathrm{eFI}_{\mathfrak E}|\widetilde{\bf U}_{\mathfrak D}$ when
$m>1$, we could partition matrix \eqref{GPpredeFI} to $D\times D$
cells with each cell an $m\times m$ matrix.
For each cell, we only take the diagonal entries of such an $m\times m$ matrix and in
this way we obtain $m$ of $D\times D$ predicted empirical Fisher
information matrices.}
\begin{equation}\label{GPpredeFI}
\E[d{\bf U}_{\mathfrak E}|\widetilde{\bf U}_{\mathfrak D}]{\bf J}_N\tp{\E[d{\bf U}_{\mathfrak E}|\widetilde{\bf U}_{\mathfrak D}]}= \widetilde{\bf L}_{\mathfrak E^1} \widetilde{\bf U}_{\mathfrak D} {\bf J}_N \tp{\widetilde{\bf U}_{\mathfrak D}} \tp{\widetilde{\bf L}_{\mathfrak E^1}} =: \widetilde{\bf L}_{\mathfrak E^1} \mathrm{gFI}_{\mathfrak D} \tp{\widetilde{\bf L}_{\mathfrak E^1}}
\end{equation}
where we name $\mathrm{gFI}_{\mathfrak D}:=\widetilde{\bf U}_{\mathfrak D} {\bf J}_N
\tp{\widetilde{\bf U}_{\mathfrak D}}$ as \emph{generalized (empirical)
Fisher information}. It is an $\tilde n\times \tilde n$ matrix that can be
pre-calculated and stored.
The left panel of Figure \ref{fig:emu} show the effect of different numbers of
design points on the emulation of empirical Fisher information, and the right panel
illustrates the emulated Fisher information on a grid mesh.
Similar results as proposition \ref{neff} exists for emulated Fisher information but we omit them here.

Note $\mathrm{gFI}_{\mathfrak D}$ in the emulated Fisher information only consists of
up to first order derivatives on design points. This avoids emulating third order
derivatives $d^3U(\vect\theta^*)$, which otherwise is inevitable in the emulation of Christoffel symbols forming the third-order tensor appearing in manifold methods.
Considering its symmetry, it seems more natural to work with the emulated Fisher information.
Denote the $(i,j)$-th element of $\mathrm{eFI}(\vect\theta^*|\mathfrak{De})$ as
$g^*_{ij}$. We have
\begin{equation}\label{deFI}
\begin{split}
g^*_{ij,k} :=& \frac{\pa g^*_{ij}}{\pa \theta^*_k} = \mathrm D_{ik}^2 U \tp{\mathrm D_j U} + \mathrm D_i U \tp{\mathrm D_{jk}^2 U} - \frac{1}{N}(\nabla_{ik}^2 U\nabla_j U + \nabla_i U \nabla_{jk}^2 U)\\
\widetilde\Gamma_{ij,k} =& \frac{1}{2}(g^*_{kj,i}+g^*_{ik,j}-g^*_{ij,k})=\mathrm D^2_{ij}U \tp{\mathrm D_k U} - \frac{1}{N}\nabla^2_{ij}U \nabla_k U = \mathrm D^2_{ij}U {\bf J}_N \tp{\mathrm D_k U}
\end{split}
\end{equation}
where $\mathrm D^2:=\nabla\otimes\mathrm D$, $\nabla^2:=\nabla\otimes\nabla$.
By a similar argument, $\widetilde{\vect\Gamma}_{\mathfrak E}|\widetilde{\bf
U}_{\mathfrak D}$ can be emulated with $\widetilde{\bf L}_{\mathfrak E^2} \mathrm{gFI}_{\mathfrak D} \tp{\widetilde{\bf L}_{\mathfrak E^1}}$,
an $mD^2\times nD$ matrix.

We then use the emulated gradients of potential energy in HMC, emulated Fisher
information and Christoffel symbols in RHMC/LMC and conduct the Metropolis acceptance
test with exactly computed potential energies. We name the corresponding algorithms as
\emph{Gaussian Process emulated (GPe)} HMC/RHMC/LMC, denoted as GPeGMC. 
In this paper, we only aim to reduce the computational cost of these expensive 
geometric quantities in HMC based algorithms.
Note in large scale data problems ($N\gg 1$), GP emulation brings computation of gradients and metrics from
$\mathcal O(ND)$ down to $\mathcal O(\tilde nD)$ for HMC, from $\mathcal O(N(f+3)D^3)$ down to $\mathcal O(\tilde n(f+1)D^3)$
for RHMC ($f$ the number of fixed point iterations) and from $\mathcal O(4ND^3)$ down to $\mathcal O(4\tilde nD^3)$ for LMC.
But at the end of each iteration, the acceptance test is done with the exact energy, which is $\mathcal O(N)$.
For large scale complicated models, the computational complexity depends on the simulator (e.g. related to mesh size of the solver in inverse problems),
regardless of the  $\tilde n$ times model simulation required at design points, GPeGMC methods requires only one simulation at the acceptance test of each iteration, the same as RWM does.
One can push further on computational economy by using the emulated potential
energies in the Metropolis acceptance probability and quantifying the bias caused by
such an inexact acceptance test \citep{welling11,korattikara13,conrad14}, e.g. after a reasonable design set is obtained using the following adaption scheme,
but it goes beyond the scope of this paper.


\section{Auto-Refinement: Online Design Pool Adaptation}\label{sec:adaption}
As mentioned previously, design points play a crucial role in emulation in general and with
GP emulation in particular. First, the size of the pool of design points should be
controlled for computational efficiency. Second, the quality of design
points directly determines the precision of the associated emulator. An ideal set of 
design points should evenly spread over the density contours so
that the GP conditioning on it models the target probability distribution well, however it is often challenging
to do so at the beginning of such an analysis, especially for computation intensive models.
Therefore, it is natural to think of adapting the initial design pool and the emulator while generating the Markov chain.

However, doing such adaptation infinitely often will disturb
the stationary distribution of the chain \citep{gelfand94,gilks98}. We refer to \emph{regeneration}
\citep{nummelin84,mykland95,gilks98,brockwell05} to allow the online adaption to occur at
certain 'regeneration times'.
Informally, a regenerative process ``restarts'' probabilistically at a set of times, called \emph{regeneration times}
\citep{brockwell05}. When the chain regenerates, the transition mechanism can be modified based on the entire history of
the chain up to that point without disturbing the consistency of MCMC estimators. 

Traditional space-filling algorithms like Latin-Hypercube, max-min etc.\citep{mckay79,morris95}
do not generate a good design set in general because rather than in some regular shape, 
the design space, not known a priori, should be characterized by the target distribution.
We therefore borrow the idea from sequential experimental design to adapt the design pool
to the shape of a target density and refine the corresponding emulator.
Starting from some initial design pool of small size, the experimental design
method can grow the pool out of some candidate set to a desirable size according to some information
criterion. Considering the setting of MCMC, it is natural to grow the design
pool by selecting candidates from previous samples which retain partial knowledge of the geometry. 
On one hand, the MCMC sampler feeds the emulator with useful candidate points to refine with; on the
other hand, the GP emulator returns geometric information (emulated gradients/metrics) efficiently for the MCMC sampler to further explore the parameter
space. The two form a mutual learning system so that they can learn from each other and gradually improve each other.

\subsection{Identifying Regeneration Times}
The main idea behind finding regeneration times is to regard the transition kernel $T(\vect\theta_{t+1}|\vect\theta_{t})$ as a mixture of two kernels, $Q$ and $R$ \cite{nummelin84,ahn13},
\begin{equation*}
T(\vect\theta_{t+1}|\vect\theta_{t})= S(\vect\theta_{t})Q(\vect\theta_{t+1}) +
(1-S(\vect\theta_{t}))R(\vect\theta_{t+1}|\vect\theta_{t})
\end{equation*}
where $Q({\vect\theta}_{t+1})$ is an \emph{independence kernel}, and the \emph{residual kernel} $R(\vect\theta_{t+1}|\vect\theta_{t})$ is defined
as follows:
\begin{equation*}
\!\!R(\vect\theta_{t+1}|\vect\theta_{t})\!\!=\!\!
\begin{dcases}
\frac{T(\vect\theta_{t+1}|\vect\theta_{t})-S(\vect\theta_{t})Q(\vect\theta_{t+1})}{1-S(\vect\theta_{t})}, &\!\!\!\!
S(\vect\theta_{t})\in [0,1)\\
1, &\!\!\!\! S(\vect\theta_{t})=1
\end{dcases}
\end{equation*}
$S(\vect\theta_{t})$ is the mixing coefficient between the two kernels such that
\begin{equation}\label{Kerineq}
T(\vect\theta_{t+1}|\vect\theta_{t}) \ge
S(\vect\theta_{t})Q(\vect\theta_{t+1}), \forall \vect\theta_{t}, \vect\theta_{t+1}
\end{equation}

Now suppose that at iteration $t$, the current state is $\vect\theta_{t}$. 
To implement this approach, we first generate $\vect\theta_{t+1}$ according to the original transition kernel
$\vect\theta_{t+1}|\vect\theta_{t}\sim T(\cdot|\vect\theta_{t})$. Then, we sample $B_{t+1}$ from a Bernoulli distribution with
probability
\begin{equation}\label{retroProb}
r(\vect\theta_{t}, \vect\theta_{t+1}) = \frac{S(\vect\theta_{t})Q(\vect\theta_{t+1})}{T(\vect\theta_{t+1}|\vect\theta_{t})}
\end{equation}
If $B_{t+1} = 1$, a regeneration has occurred, then we discard $\vect\theta_{t+1}$ and sample it from the independence kernel
$\vect\theta_{t+1}\sim Q(\cdot)$.
At regeneration times, we refine design points and the associated emulator using
the past sample path.

Ideally, we would like to evaluate regeneration times in terms of the HMC style
transition kernel. In general, however, this is quite difficult for such a
Metropolis algorithm. On the other hand, regenerations are easily achieved for
the independence sampler (i.e., the proposed state is independent from the
current state) as long as the proposal distribution is close to the target
distribution \citep{gilks98}. Therefore, we can specify a hybrid sampler that
consists of the original proposal distribution (GPeGMC) and the
independence sampler, and adapt both proposal distributions whenever a
regeneration is obtained on an independence sampler step \citep{gilks98}. In our
method, we systematically alternate between GPeGMC and the independence
sampler while evaluating regeneration times based on the independence sampler
only \citep{ahn13,lan14b}.

\begin{figure}[t]
  \begin{center}
    \includegraphics[width=.9\textwidth,height=.4\textwidth]{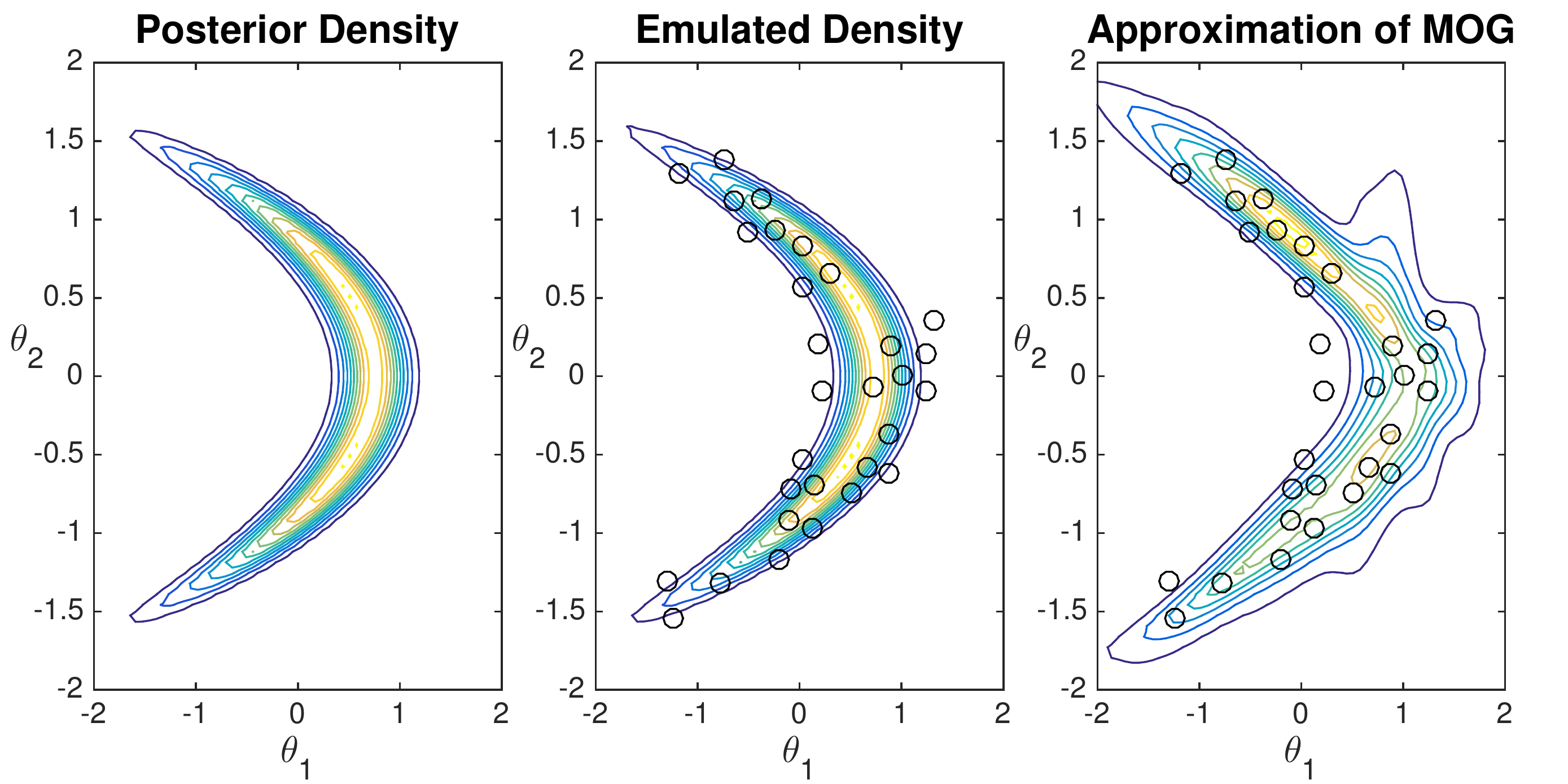}
  \end{center}
  \caption{Approximations to the true density. Left: true posterior density;
  Middle: posterior density approximated by Gaussian Process with 30 design
  points; Right: approximate density given by mixture of Gaussians centered at
  the same design points.}
  \label{fig:approxtest}
\end{figure}

More specifically, $T(\vect\theta_{t+1}|\vect\theta_{t})$, $S(\vect\theta_{t})$ and $Q(\vect\theta_{t+1})$ are defined as
follows to satisfy \eqref{Kerineq}:
\begin{align}
T(\vect\theta_{t+1}|\vect\theta_{t}) &= q(\vect\theta_{t+1})
\min\left\{1,\frac{\pi(\vect\theta_{t+1})/q(\vect\theta_{t+1})}{\pi(\vect\theta_{t})/q(\vect\theta_{t})}\right\}
\label{TSQ:T}\\
S(\vect\theta_{t}) &= \min\left\{1,\frac{c}{\pi(\vect\theta_{t})/q(\vect\theta_{t})}\right\} \label{TSQ:S}\\
Q(\vect\theta_{t+1}) & = q(\vect\theta_{t+1}) \min\left\{1,\frac{\pi(\vect\theta_{t+1})/q(\vect\theta_{t+1})}{c}\right\}
\label{TSQ:Q}
\end{align}
A natural choice for $q(\cdot)$ could be based on the emulated potential
energy as we have seen in figure \ref{fig:approxtest} that it approximates the target distribution very well:
\begin{equation}
\begin{split}
q(\vect\theta^*) \propto & \exp(-\E[U(\vect\theta^*|\mathfrak{De})]) = \exp\left\{-{\bf h}(\vect\theta^*)\widehat{\vect\beta}-{\bf C}(\vect\theta^*,\widetilde{\mathfrak D})\widetilde{\vect\gamma}\right\}\\
\propto & \exp\left\{-\tp{\widetilde{\vect\beta}}\vect\theta^*-\sum_{j=1}^n\left[\tilde\gamma_j+\sum_{k=1}^D 2\rho_k(\theta^*_k-\theta^j_k)\tilde\gamma_{kn+j}\right]\cdot\right.\\
&\phantom{\exp\left\{-\tp{\widetilde{\vect\beta}}\vect\theta^*-\sum_{j=1}^n\right.} \left.\exp\left[-\tp{(\vect\theta^*-\vect\theta^j)}\mathrm{diag}(\vect\rho)(\vect\theta^*-\vect\theta^j)\right]\right\}
\end{split}
\end{equation}
where $\widetilde\gamma=\widetilde{\bf C}_{\mathfrak D}^{-1}(\tilde{\bf u}_{\mathfrak D}-\widetilde{\bf H}_{\mathfrak D}\widehat{\vect\beta})$.
However in general it is very difficult to directly sample from a distribution with such
density. In practice, we find that it works well to set $q(\cdot)$ to be a density comprised of, for example, a mixture of Gaussians centered at the design
design points $\mathfrak{De}$ with empirical Fisher matrices
$\mathrm{eFI}(\mathfrak{De})$ as the precision matrices and posterior
probabilities $\exp(-U(\mathfrak{De}))$ as their relative weights.
Figure \ref{fig:approxtest} shows that with a reasonably good design, such a
mixture of Gaussians could approximate the true posterior density reasonably well for practical purposes.

\subsection{Refining the design set}
When the Markov Chain regenerates, we refer to an experimental design method based on
\emph{mutual information} to refine the design set and update the associated GP emulator accordingly.
Such a method sequentially selects design points from some candidate set formed by previous MCMC samples.
A design point is chosen by optimizing the mutual information gain of adding a point to the design pool.

In classical information theory, mutual information \citep{mackay03} is a
standard measure that has been successfully applied to sensor network design
\citep{caselton84,guestrin05}, experimental design \citep{huan13}, and
optimization \citep{contal13}. Consider two random vectors $U$ and $U'$ with
marginal pdfs $p_{U}({\bf u})$ and $p_{U'}({\bf u}')$, and joint pdf
$p_{UU'}({\bf u},{\bf u}')$. The \emph{mutual information} between them, denoted
by $I(U;U')$ is equivalent to the Kullback-Leibler divergence
$D_{KL}(\cdot||\cdot)$ between $p_{UU'}$ and $p_{U}p_{U'}$ and linked to the
entropy \citep{mackay03}:
\begin{equation}\label{MI}
I(U;U')=D_{KL}(p_{UU'}||p_{U}p_{U'})=H(U)-H(U|U')
\end{equation}
where the entropy in the Gaussian Process setting is as follows
\begin{equation}\label{Ent}
H(U(\mathfrak{De})) \propto \frac{1}{2}\log \det {\bf C}_{\mathfrak D}
\end{equation}

Given a current design $\mathfrak{De}$ and a candidate set $\vect\Theta_{cand}$,
to choose the most informative subset from $\vect\Theta_{cand}$ to add to $\mathfrak{De}$, \cite{guestrin05} propose a
greedy algorithm based on the following sequential optimization of mutual
information:
\begin{equation}\label{greedyMI}
\begin{split}
\vect\theta^*&=\arg\max_{\vect\theta\in\vect\Theta_{cand}} I(U(\mathfrak{De}\cup\{\vect\theta\});U(\mathfrak{De}^c\backslash\{\vect\theta\}))-I(U(\mathfrak{De});U(\mathfrak{De}^c))\\
&=\arg\max_{\vect\theta\in\vect\Theta_{cand}} H(U(\vect\theta)|{\bf u}_{\mathfrak D})-H(U(\vect\theta)|U(\mathfrak{De}^c\backslash\{\vect\theta\}))\\
&=\arg\max_{\vect\theta\in\vect\Theta_{cand}} \V(U(\vect\theta)|{\bf u}_{\mathfrak D})/\V(U(\vect\theta)|U(\mathfrak{De}^c\backslash\{\vect\theta\}))
\end{split}
\end{equation}
where $\mathfrak{De}\cup\mathfrak{De}^c=\vect\Theta$ forms the whole design space.
Based on the same idea, \cite{beck14} improves it and adopts it for computer
experiments. Therefore, the algorithm is named \emph{Mutual Information for Computer
Experiments (MICE)} \citep{beck14} and is summarized in algorithm \ref{Alg:MICE}.

We now adapt MICE \citep{beck14} for our specific purpose of refining the MCMC
transition kernel.
Starting with the current design set, we select a small
number, $k$, of points from it as our \emph{initial} design, and
combine the rest of the design points with points collected from the previous
regeneration tour, to form a candidate set.
There are two advantages of using MCMC samples for candidates in MICE.
One is that MCMC samples retain partial knowledge of the geometry of the target distribution
therefore are more informative than some random points. The other is that, potential energies at these points,
required in MICE, have already been calculated at the Metropolis test step and can be recycled.

First, before running MICE, we can pre-process the candidate set to narrow it down
to a smaller subset of points controlling their pairwise distance by a max-min method.
This is recommended not only to save computation of the optimization \eqref{greedyMI},
but also to avoid poor conditioning of the covariance in the GP.
Second, for the sake of computational efficiency, instead of continuously growing
the design size, we 'refresh' the design set with a flexible size to be determined by the algorithm.
More specifically, after pre-processing the candidates, we use MICE to select points one by one 
from the processed candidate set, and add it to the initial design set until this process is stopped 
by some criterion, e.g., MSPE falls below some threshold.
Third, we do such adaptation in a more structured way. Regeneration is tested
with some interval (e.g. every 20 iterations) because too short a regeneration tour
will not provide many informative candidate points, and the adaptation is stopped
when we reach a satisfactory (e.g. by testing MSPE on a set of randomly selected samples,
or monitoring the entropy to reach its stationarity) design set.

We summarize the adaptive GP emulated GMC (adpGPeGMC) in algorithm \ref{Alg:adpGPeGMC}.
Figure \ref{fig:designevolsnap} shows how the design pool is evolved by the MICE algorithm.
Even starting with a bad design (crowed at the starting point), modified
MICE can gradually spread them over the target density contour until it reaches the final design based on which
Gaussian process accurately emulates the true probability distribution.

\begin{figure}[t]
  \begin{center}
    \includegraphics[width=1\textwidth,height=.4\textwidth]{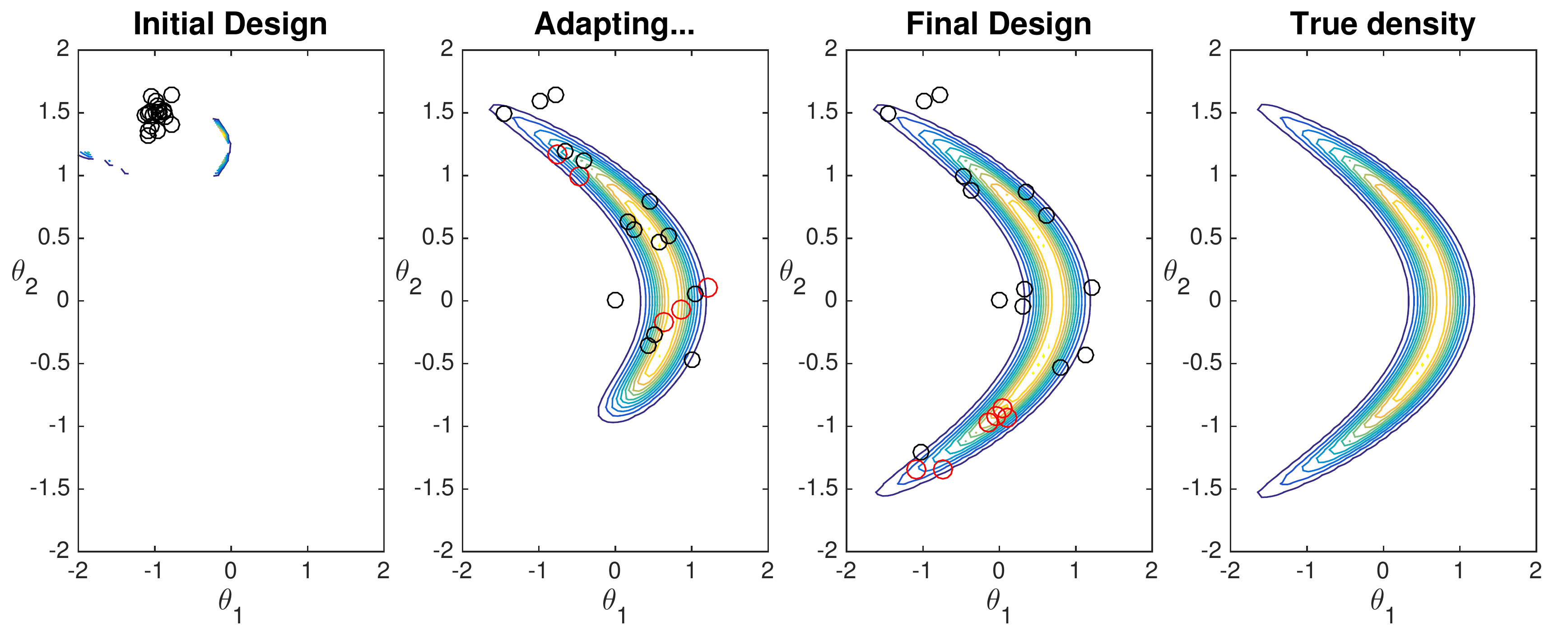}
  \end{center}
  \caption{The evolution of the design pool by the MICE algorithm. Black circles are
  design points and red circles are new design points added to the
  design in each adaptation. All the density contours are produced by GP
  emulation based on the current design except the true one.}
  \label{fig:designevolsnap}
\end{figure}


\section{Experimental Evaluation}
In this section, we use two synthetic examples and one oil reservoir problem to evaluate our GPeGMC, compared to the full version of these geometric Monte Carlo methods. We use a time-normalized effective sample size (ESS) to compare these methods \citep{girolami11}. For $B$ posterior samples, the ESS for each parameter is defined as  $\textrm{ESS} =B[1 + 2\Sigma_{k=1}^{K}\gamma(k)]^{-1}$, where $\Sigma_{k=1}^{K}\gamma(k)$ is the sum of $K$ monotone sample autocorrelations \citep{geyer92}. We use the minimum ESS normalized by CPU time (s), $\min(\textrm{ESS})/\textrm{s}$, as the measure of sampling efficiency.
We also use Relative Error of Mean (REM) and Relative Error of Covariance (REC) to measure how fast different algorithms reduce the error in estimating mean and covariance of parameters given the limited computation budget. They are defined as $\Vert \bar{\vect\theta(t)}-\E\vect\theta\Vert_2/ \Vert \E\vect\theta\Vert_2$ and $\Vert s(\vect\theta(t))-\Cov\vect\theta\Vert_2/ \Vert \Cov\vect\theta\Vert_2$ respectively, where $\vect\theta(t)$ means samples collected up to time $t$.
All computer codes and data sets discussed in this paper are publicly available at \url{http://warwick.ac.uk/slan}.

\begin{figure}[t]
  \begin{center}
    \includegraphics[width=.95\textwidth,height=.35\textwidth]{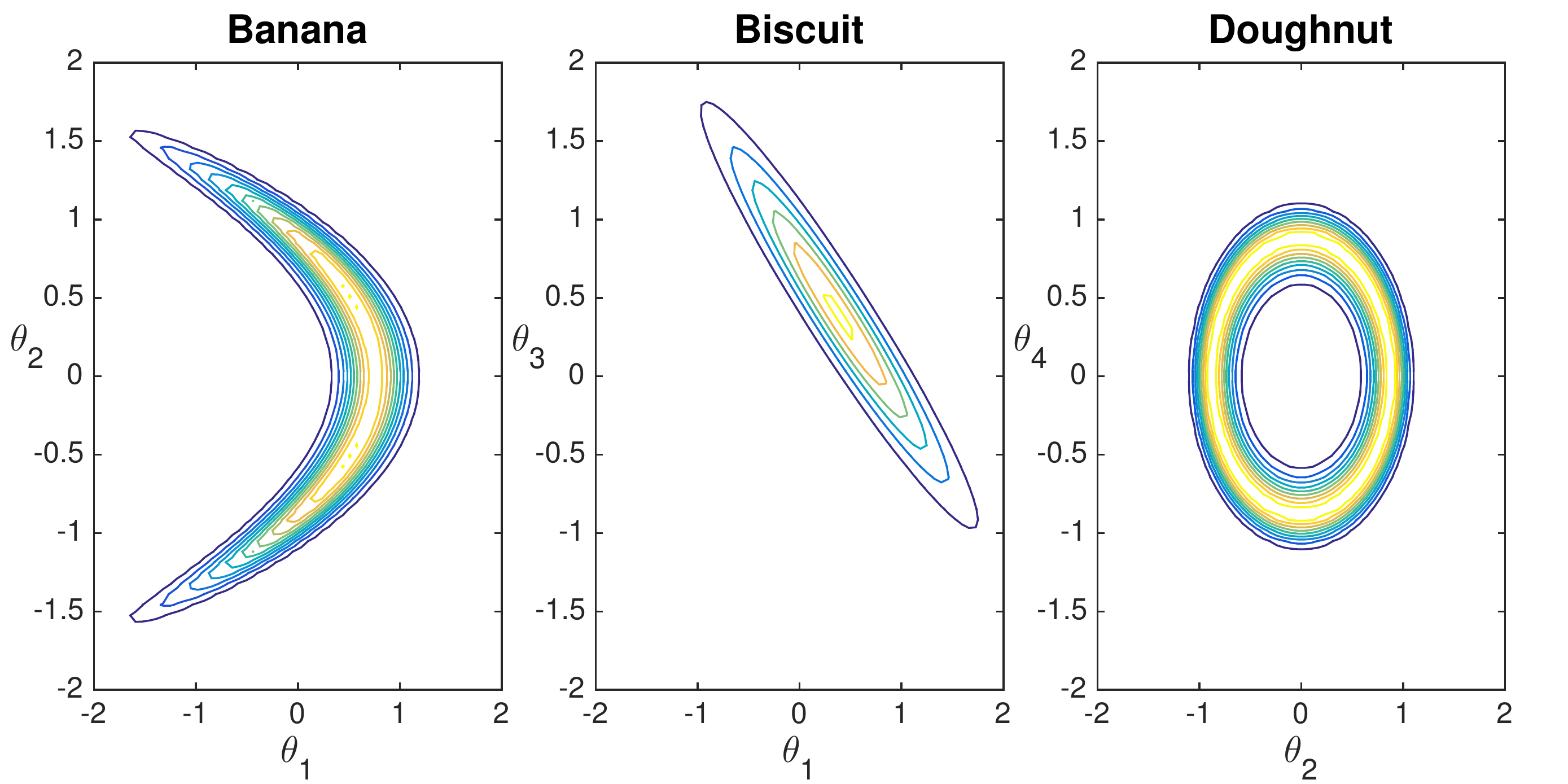}
  \end{center}
  \caption{Banana-Biscuit-Doughnut distribution: generated with $N=100$, $\mu_y=1,
  \sigma_{y}=2$, and $\sigma_{\theta}=1$ as in \eqref{bbd}.}
  \label{fig:bbd}
\end{figure}

\subsection{Simulation: Banana-Biscuit-Doughnut distribution}
Let us first investigate the generalisation of the previously discussed 2d Banana
shaped distribution:
\begin{equation}\label{bbd}
\begin{aligned}
y|\vect\theta & \sim \mathcal N(\mu_y,\sigma^{2}_y),\quad \mu_y:=\sum_{k=1}^{\lceil D/2\rceil}\theta_{2k-1}+\sum_{k=1}^{\lfloor D/2\rfloor}\theta_{2k}^{2}\\
\theta_i & \overset{iid}{\sim} \mathcal N(0,\sigma^{2}_{\theta})
\end{aligned}
\end{equation}
The data $\{y_n\}_{n=1}^N$ are generated with chosen $\mu_y, \sigma_{y}$, and $\sigma_{\theta}$.
If we consider $D=4$, then the posterior $\vect\theta|\{y_n\}$ looks like a
banana in $(1,2)$ dimension, a biscuit in $(1,3)$ dimension and a doughnut in
$(2,4)$ dimension. Therefore we name the distribution as 'Banana-Biscuit-Doughnut (BBD)'
distribution as depicted in figure \ref{fig:bbd}.

This BBD distribution is complex as three different shapes are twisted around the
origin. To make it more data intensive, we generate $N=3\times10^6$ data $y_n$
with $\mu_y=0,\sigma_{y}=10^4$, and $\sigma_{\theta}=1$. Now we apply all the
geometric Monte Carlo methods to sample $\vect\theta|\{y_n\}$ and compare their
full versions with the GP emulated versions in terms of sampling efficiency in table \ref{tab:bbd}.
Results are summarised for 10000 samples after burn-in.
\begin{table}[ht]
\centering
\begin{tabular}{l|ccccc}
  \hline
Algorithm & AP & s/iter & ESS & minESS/s & spdup \\ 
  \hline
RWM & 0.68 & 9.84E-03 & (3,4,8) & 0.03 & 1.00 \\ 
  HMC & 0.84 & 8.28E-02 & (552,873,1102) & 0.67 & 21.97 \\ 
  GPeHMC & 0.78 & 2.84E-02 & (622,653,754) & 2.19 & 72.14 \\
  RHMC & -- & -- & -- & -- & -- \\ 
  GPeRHMC & 0.76 & 1.32E-01 & (319,560,756) & 0.24 & 7.95 \\
  LMC & 0.91 & 2.21E-00 & (910,1120,1251) & 0.04 & 1.36 \\ 
  GPeLMC & 0.78 & 2.24E-02 & (574,609,823) & 2.56 & 84.36 \\ 
  adpGPeLMC & 0.67 & 2.38E-02 & (252,322,509) & 1.06 & 34.90 \\ 
   \hline
\end{tabular}
\caption{Sampling Efficiency in the BBD distribution. AP is the acceptance probability, s/iter is the CPU time (second) for each iteration, ESS has (min., med., max.) and minESS/s is the time-normalized ESS. Spdup is the speed up of sampling efficiency measured by minESS/s with RWM as the baseline.} 
\label{tab:bbd}
\end{table}

In this example, 
it is extremely time consuming to scan $3\times 10^6$ items for each gradient evaluation.
Though the metric tensor, expected Fisher information, can be explicitly calculated, for illustrative purposes
we use the empirical Fisher information which estimates the expected Fisher information with gradients as a means to compare in a fair way the
emulated geometric Monte Carlo methods to their full versions.
For each GPeGMC, 40 design points (samples) are chosen from a long run HMC using the MICE algorithm. AdpGPeLMC
is the GP emulated LMC with online adaptation discussed in section \ref{sec:adaption}.

The benefit of higher ESS by using the metric in LMC is completely offset by the computational cost of geometric quantities, 
making LMC close to RWM in efficiency after normalizing total sampling time. RHMC is even more time consuming due to
the repeated metric evaluation in the implicit steps of the generalised leapfrog integrator, so we exclude it from the comparison.
Compared with those original geometric Monte Carlo methods,
our proposed emulated methods in general yield less raw ESS due to the GP emulation,
but at substantially lower computational cost, yielding much more efficient algorithms.
However, due to the complexity of the distribution, online adaptation requires some time to find an appropriate configuration that captures the geometric features, and as such
we do not observe any advantage of adpGPeLMC over GPeLMC in this example, though adpGPeLMC is still much more efficient than LMC.

\begin{figure}[t]
  \begin{center}
    \includegraphics[width=.95\textwidth,height=.75\textwidth]{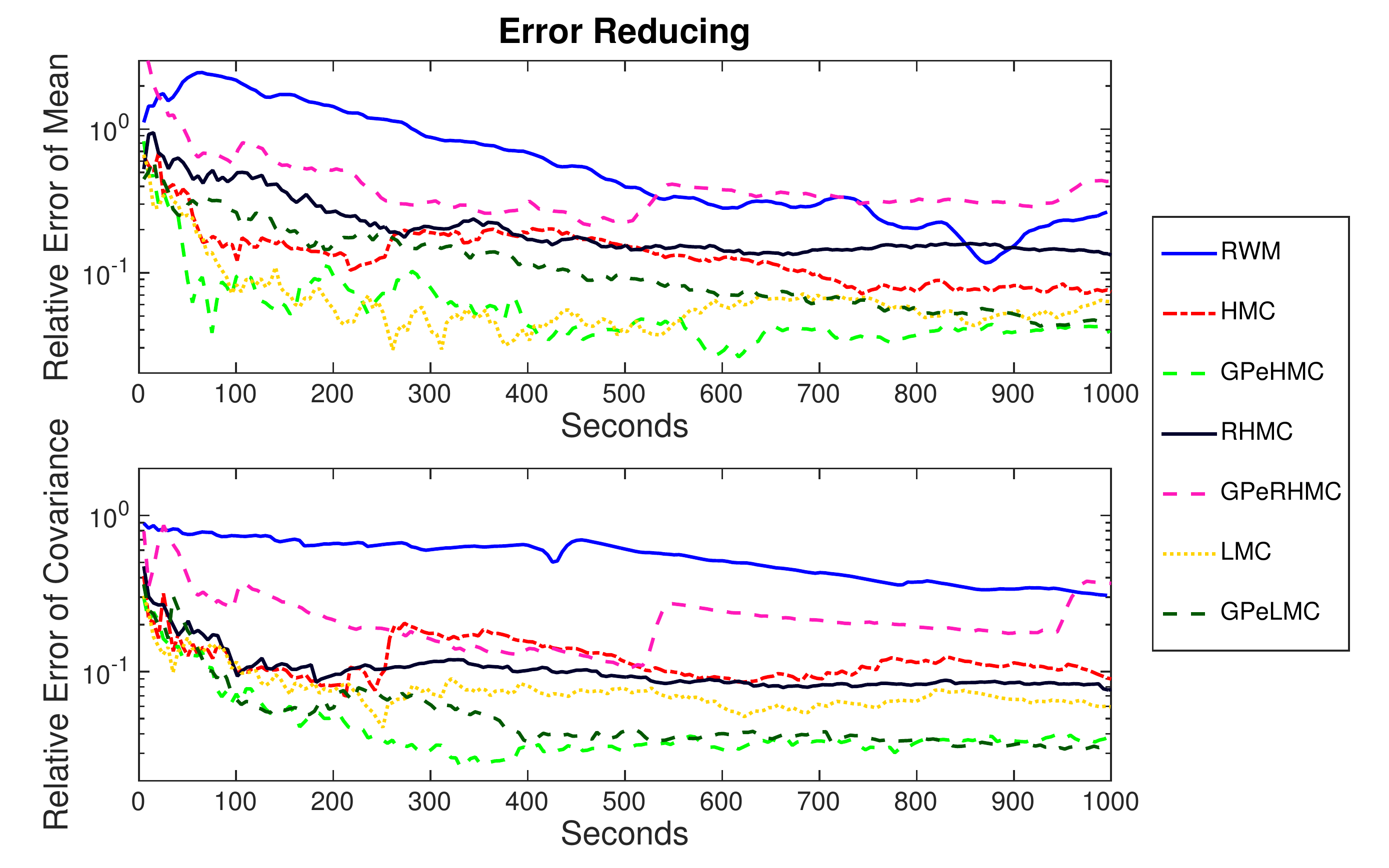}
  \end{center}
  \caption{Relative error of mean (left) and relative error of covariance
  (right) in sampling from BBD distribution.}
  \label{fig:errbbd}
\end{figure}

Since GP emulated Monte Carlo methods generally take less time to generate a sample,
they are usually fast in reducing errors of point estimates, eg. mean and covariance, as shown in figure \ref{fig:errbbd},
where analytical expression of the expected Fisher information is used.
This is important in obtaining good estimates quickly given a limited computational budget.


\subsection{Elliptic PDE}
Now we consider a canonical inverse problem involving inference of the diffusion
coefficients in the following elliptic PDE \citep{dashti11,conrad14} defined on the unit
square $[0,1]^2$:
\begin{equation}\label{ellipticPDE}
\begin{aligned}
\nabla_{\bf x}\cdot (c({\bf x},\vect\theta)\nabla_{\bf x}u({\bf x},\vect\theta)) &= 0\\
u({\bf x},\vect\theta)|_{x_2=0} & = x_1\\
u({\bf x},\vect\theta)|_{x_2=1} & = 1-x_1\\
\left.\frac{\pa u({\bf x},\vect\theta)}{\pa x_1}\right|_{x_1=0} &= \left.\frac{\pa u({\bf x},\vect\theta)}{\pa x_1}\right|_{x_1=1} = 0
\end{aligned}
\end{equation}
This PDE serves as a simple model of steady-state flow in aquifers and other
subsurface systems; $c$ can represent the permeability of a porous medium while
$u$ represents the hydraulic head.

The observations $\{y_i\}$ arise from the solutions on a $11\times 11$ grid
contaminated by additive Gaussian error $\eps_j\sim \mathcal N(0,0.1^2)$:
\begin{equation*}
y_i = u({\bf x}_i,\vect\theta) + \eps_i
\end{equation*}
Endow the diffusive field $c({\bf x})$ with a log-Gaussian process prior, we want to
infer its posterior. This prior allows the field to be approximated by a
Karhunen-Lo\`eve (K-L) expansion \citep{adler81}:
\begin{equation*}
c({\bf x},\vect\theta) \approx
\exp\left(\sum_{d=1}^D\theta_d\sqrt{\lambda_d}c_d({\bf x})\right)
\end{equation*}
where $\lambda_d$ and $c_d({\bf x})$ are the eigenvalues and eigenfunctions of
integral operator on $[0,1]^2$ defined by the Gaussian kernel. The parameters
$\vect\theta$ are endowed with independent Gaussian priors $\theta_i\sim\mathcal
N(0,1)$. Here, the K-L expansion is truncated at $D=6$.

The statistical model is relatively simple. However, each likelihood evaluation
involves numerically solving a forward PDE based on a $20\times 20$ regular
mesh, which could be computationally intensive. Geometric MCMC methods require further
derivatives of likelihood, which involve partial derivatives $\left\{\frac{\pa
u}{\pa\theta_d}\right\}$, which are even more expensive to calculate.
We apply our emulated algorithms to this problem and compare
the sampling efficiency of different algorithms.
The result is summarised in table \ref{tab:ellipticPDE}.
\begin{table}[ht]
\centering
\begin{tabular}{l|ccccc}
  \hline
Algorithm & AP & s/iter & ESS & minESS/s & spdup \\ 
  \hline
RWM & 0.57 & 2.70E-02 & (69,94,249) & 0.25 & 1.00 \\ 
  HMC & 0.76 & 4.35E-01 & (3169,4357,5082) & 0.73 & 2.87 \\ 
  GPeHMC & 0.57 & 1.56E-02 & (609,1328,2265) & 3.91 & 15.38 \\ 
  RHMC & 0.87 & 2.22E+00 & (5073,5802,6485) & 0.23 & 0.90 \\ 
  GPeRHMC & 0.65 & 7.35E-02 & (614,1224,1457) & 0.83 & 3.29 \\ 
  LMC & 0.72 & 3.92E-01 & (5170,5804,6214) & 1.32 & 5.19 \\ 
  GPeLMC & 0.64 & 2.65E-02 & (774,1427,1754) & 2.92 & 11.48 \\ 
  adpGPeLMC & 0.94 & 8.71E-02 & (3328,4058,4543) & 3.82 & 15.03 \\ 
   \hline
\end{tabular}
\caption{Sampling Efficiency in Elliptic PDE. AP is the acceptance probability, s/iter is the CPU time (second) for each iteration, ESS has (min., med., max.) and minESS/s is the time-normalized ESS. Spdup is the speed up of sampling efficiency measured by minESS/s with RWM as the baseline.} 
\label{tab:ellipticPDE}
\end{table}

\begin{figure}[t]
  \begin{center}
    \includegraphics[width=.95\textwidth,height=.75\textwidth]{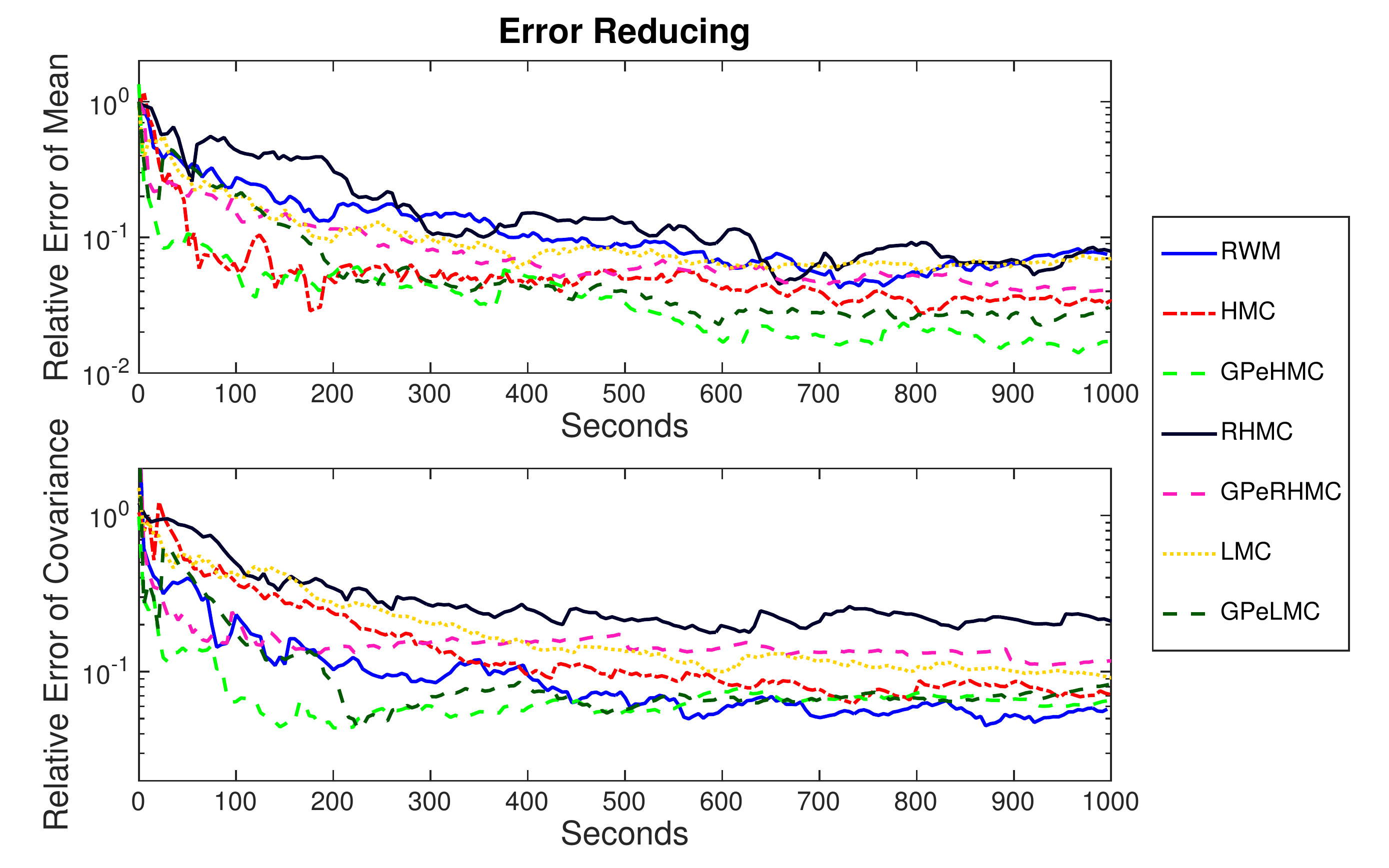}
  \end{center}
  \caption{Relative error of mean (left) and relative error of covariance
  (right) in Elliptic PDE problem.}
  \label{fig:errEllipticPDE}
\end{figure}

We run each algorithm for 15000 iterations and burn in the first 5000. We tune
the step sizes so that they have an acceptance rate around $70\%$. For HMC,
RHMC and LMC, since they require solving an elliptic PDE \eqref{ellipticPDE} for each
integration step, we parallelize the computation.
Again we observe a drop in the raw ESS when comparing emulated algorithms with the full versions,
but an increase in efficiency due to the computational time cut by GP emulation.
The pairwise posterior contours of the parameter $\vect\theta$ are included in figure \ref{fig:ellipticPDE}.
Although not completely like a Gaussian, the posterior distribution is not as complicated as the BBD distribution.
Therefore, the online adaption (in adpGPeLMC) can obtain a better configuration and exhibit further advantage
compared to pre-fixing the design set (in GPeHMC and GPeLMC) by selecting design points
from the result of HMC after a long run.
We also compare different algorithms in error reducing speed in the following
figure \ref{fig:errEllipticPDE}.

\begin{figure}[t]
  \begin{center}
    \includegraphics[width=.8\textwidth,height=.45\textwidth]{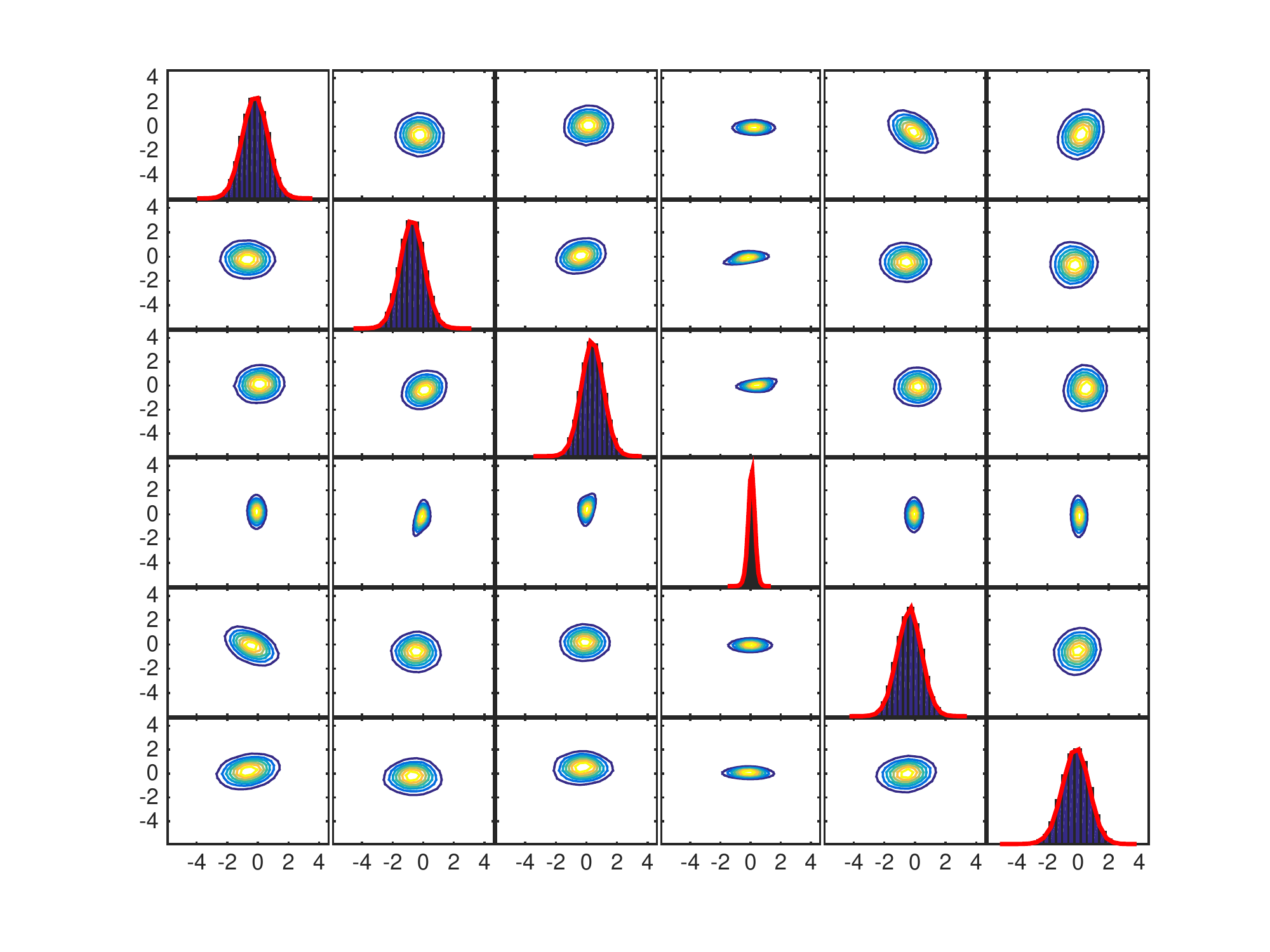}
  \end{center}
  \caption{The pairwise posterior contours of the parameter $\vect\theta$ in
  inverse problem of a classic elliptic PDE.}
  \label{fig:ellipticPDE}
\end{figure}

\subsection{Teal South oil reservoir}
\begin{figure}[t]
  \begin{center}
    \includegraphics[width=.49\textwidth,height=.4\textwidth]{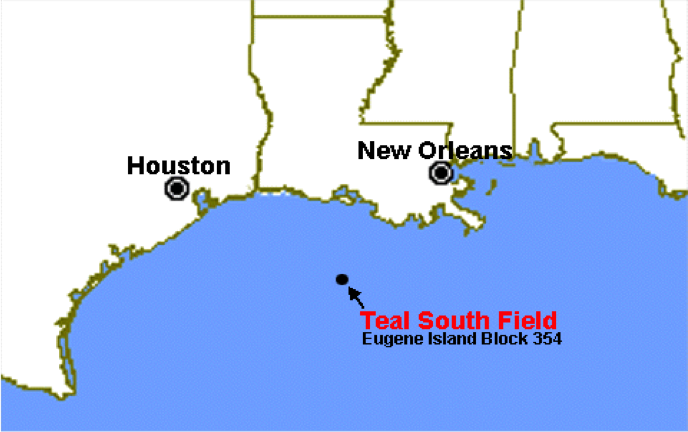}
    \includegraphics[width=.49\textwidth,height=.4\textwidth]{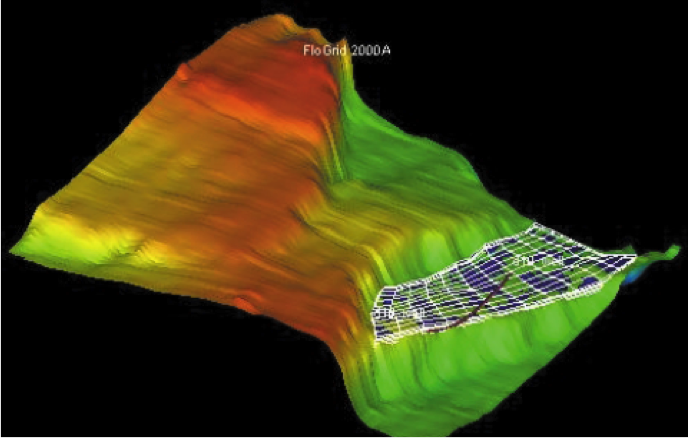}
  \end{center}
  \caption{Teal South oil field}
  \label{fig:tealsouth}
\end{figure}

Teal South (Figure \ref{fig:tealsouth}) is a small oil field located in the northern Gulf of Mexico with the information about the field in the public domain. 
The field has a single well through which oil, water and gas are produced, with monthly production rates available for each phase. 
Teal South has been the subject of a number of studies \citep{pennington01,hajizadeh11,islam14,mohamed10}. 
We used a simple reservoir model with an $11\times11\times5$ grid created by \cite{mohamed10}, with 9 unknown parameters to describe reservoir uncertainty. 
The 9 parameters are: $k_h$ (horizontal permeabilities) for each of the 5 layers of the field, $k_v/k_h$ (vertical to horizontal permeability ratio), aquifer strength, rock compressibility and porosity. 
A set of PDEs describing conservation of mass for each of the phases (oil, water, gas), along with Darcy's law are solved to simulate the field oil production rate (FOPR) for a period of 1200 days starting from November 1996. 
Given the inputs (9 parameters), the outputs (FOPR as a time series) are generated by running tNavigator \citep{RFD}, an oil reservoir simulation package.

In general, reservoir simulation is computationally demanding, with a single run of a reservoir simulation model taking anywhere from 15 minutes to several hours on high end workstations (see for example \citep{christie13}). 
Teal South runs much faster than this, as it is a simple model, but even with such a simple model we cannot afford the tens of thousands of function evaluations that would be needed for RWM. 
To start the inference process, we generated 1000 samples by stochastic optimisation algorithms (differential evolution \citep{hajizadeh10}, particle swarm optimization \citep{mohamed11}, and a Bayesian optimization algorithm \citep{abdollahzadeh12}). 
Since parameters are in different scales, we normalize these 1000 samples. The following inference is based on emulation with these 1000 design points.

One advantage of GP emulated Monte Carlo is that Hamiltonian algorithms can still be
applicable in the absence of derivative and metric information because we can emulate them.
Given weak priors, we sample from the posterior of the 9 parameters using RWM,
GPeHMC (full versions of HMC are not available
here due to the blackbox nature of the simulator codes), and compare their sampling efficiency in table \ref{tab:tealsouth}.

\begin{table}[ht]
\centering
\begin{tabular}{l|ccccc}
  \hline
Algorithm & AP & s/iter & ESS & minESS/s & spdup \\ 
  \hline
RWM & 0.77 & 1.35E-03 & (43,81,116) & 3.20 & 1.00 \\ 
  GPeHMC & 0.89 & 4.38E-03 & (5808,6023,6124) & 132.73 & 41.49 \\ 
   \hline
\end{tabular}
\caption{Sampling Efficiency in Teal South oil reservoir problem. AP is the acceptance probability, s/iter is the CPU time (second) for each iteration, ESS has (min., med., max.) and minESS/s is the time-normalized ESS. Spdup is the speed up of sampling efficiency measured by minESS/s with RWM as the baseline.} 
\label{tab:tealsouth}
\end{table}

Note, the emulated gradient apparently helps to improve the sampling efficiency, furthermore, given the approximate Gaussian form of the posterior there is no requirement for the higher order metric components of the manifold methods.
In figure \ref{fig:sampleanalysis}, the left panel compares the
samples generated by different algorithms and the right panel compares their
convergence rate. GPeHMC not only converges faster but also generates less
auto-correlated samples than the others.

\begin{figure}[t]
  \begin{center}
    \includegraphics[width=.55\textwidth,height=.65\textwidth]{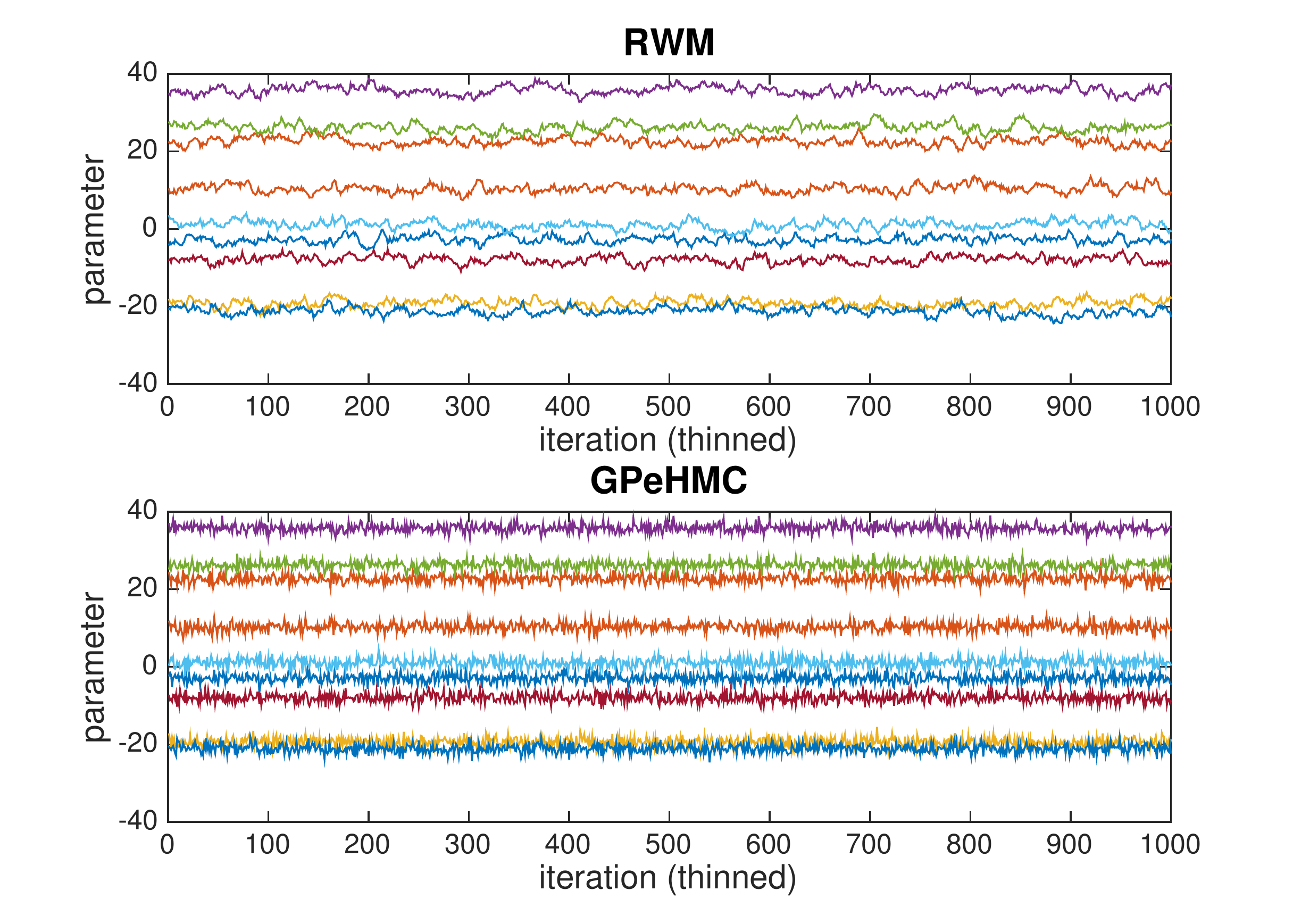}
    \includegraphics[width=.44\textwidth,height=.65\textwidth]{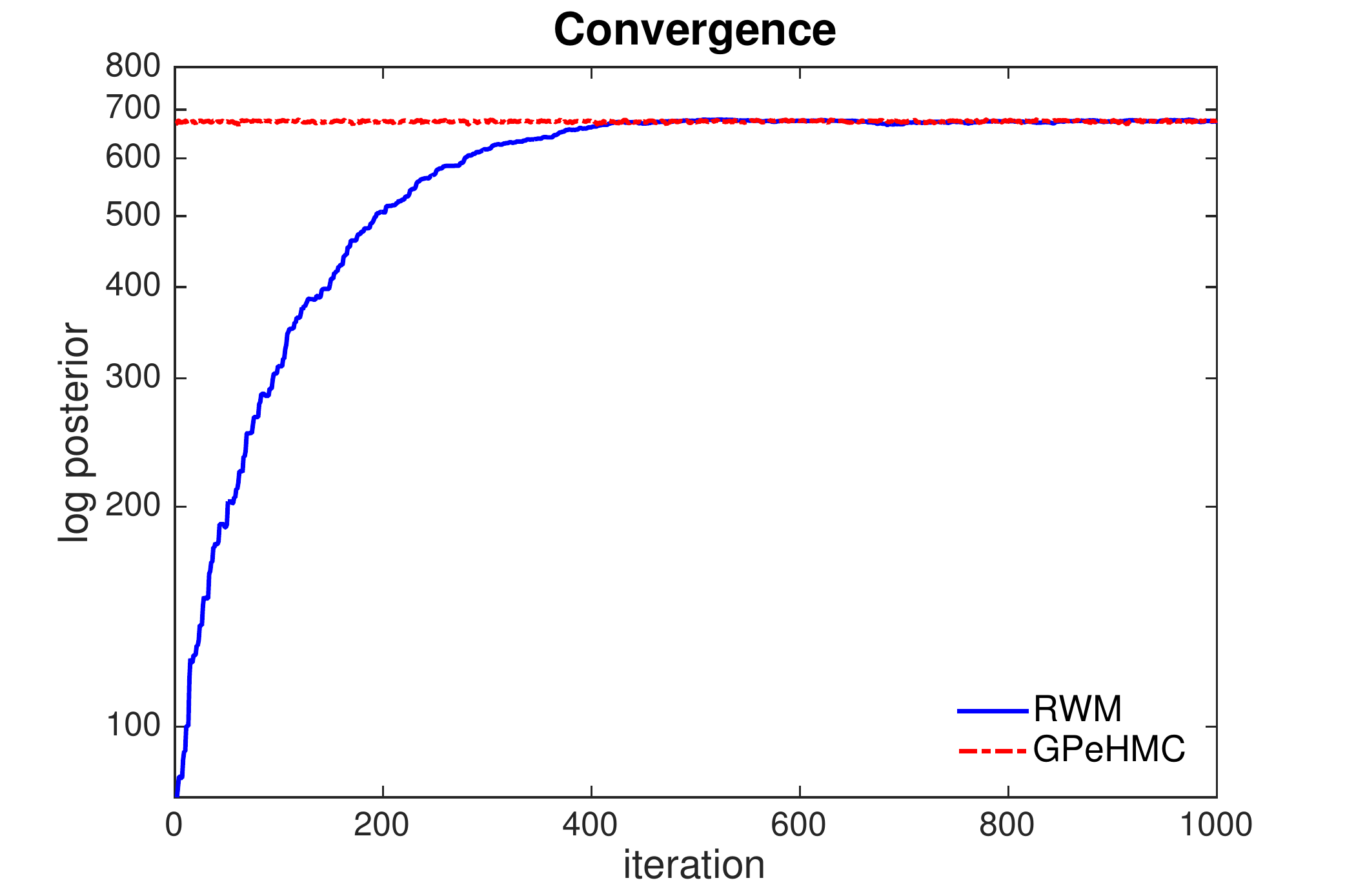}
  \end{center}
  \caption{left: Samples (thinned for 1 every 10 samples); right: log posterior
  (no burning or thinning)  in Teal South oil reservoir problem}
  \label{fig:sampleanalysis}
\end{figure}


\section{Conclusion}
Geometric information of a probability distribution improves the efficiency of MCMC samplers in
exploring the parameter space thus it can improve the mixing rate of the
resulting Markov chain. In HMC, gradient information of the distribution guides
the sampler to explore high density regions; in RHMC/LMC, metric information
further adapts the sampler to the local geometry of the distribution. They
suppress the random walk behaviour in the classical Metropolis algorithm however
impose computational challenges by the requirement of these geometric objects (gradient, metric, etc).

In this paper, we investigate emulation using Gaussian Processes as a cheaper
alternative to exact calculation of geometric information, with the aim to
reduce the computational cost of these geometric Monte Carlo methods.
It is believed to be the first attempt to emulate higher order
derivatives in MCMC samplers in order to improve their efficiency. Furthermore,
we observe the importance of design points selection for Gaussian Process emulation and
the necessity of refinement of the design set as it is impractical to initialise
a good set. This is also one of the first attempts to introduce experiment design
methods to emulation for MCMC. We refer to the regeneration technique to determine
legal times when the adaptation is allowed based on the previous history. When the
chain regenerates, the MICE algorithm is used to sequentially refine the design set
and update the associated Gaussian Process emulator.

Simulation studies and real problems have shown a substantial advantage of emulated
geometric Monte Carlo methods over their original versions in this setting. There are many
directions to further improve the proposed methods. One of them could be 
a flexible mechanism to determine proper design size depending on the
dimension and complexity of the problem. It is natural to think that a 100
dimensional distribution that is nearly Gaussian may not need more design points
than a 2d highly skew, ill-shaped distribution to provide decent emulation.
Another direction could be the local predictor using a subset of neighbouring
points instead of the whole set of design points. This is crucial to save
computational cost as dimension grows. Computation of a large correlation matrix
is involved in the design refinement step and Hierarchical matrix factorization
\citep{ho13a,ho13b} may be a direction that can contribute. Lastly, it is
natural to parallelize the Markov chains using our method as Gaussian Processes can predict on
multiple points simultaneously. This can be applied to parallel tempering \citep{swendsen86,neal96a,earl05}.

\section*{Acknowledgement}
We thank Patrick R. Conrad for assistance in running the elliptic PDE example. SL is supported by EPSRC Programme Grant, Enabling Quantification of Uncertainty in Inverse Problems (\href{http://www2.warwick.ac.uk/fac/sci/maths/research/grants/equip/}{EQUIP}), EP/K034154/1.
TB is supported by Department of Energy grants DE-SC0010518 and DE-SC0011118. MC is partially supported by EPSRC Programme Grant EQUIP, EP/K034154/1.
MG is funded by an EPSRC Established Career Research Fellowship EP/J016934/2.

\section*{References}

\bibliography{references}

\newpage
\begin{center}
{\huge \bf Appendix: Calculations and Proofs}
\end{center}
\appendix

\section{Maximum Likelihood Estimate of $\vect\rho$}\label{apdx:MLE}
The likelihood of $\vect\rho$ is as follows:
\begin{equation}\label{likrho}
L(\vect\rho) \propto (\widehat{\sigma}^2(\vect\rho))^{-(n-q)/2}|{\bf C}_{\mathfrak D}(\vect\rho)|^{-1/2}|\tp{\bf H}_{\mathfrak D}{\bf C}_{\mathfrak D}^{-1}(\vect\rho){\bf H}_{\mathfrak D}|^{-1/2}
\end{equation}
To find MLE of $\vect\rho$, we want to use Trust Region Reflective algorithm to
optimize the following function
\begin{equation}
l(\vect\rho) = -(n-q)/2\log\widehat{\sigma}^2(\vect\rho)-1/2\log\det{\bf C}_{\mathfrak D}(\vect\rho)-1/2\log\det{\bf B}_{\mathfrak D}(\vect\rho)
\end{equation}
And we need the gradients and Hessians \citep[See also][]{andrianakis09}:
\begin{equation}\label{likrho_grad}
\begin{aligned}
\frac{\pa l}{\pa\rho_d} =& -\frac{n-q}{2\widehat{\sigma}^2}\frac{\pa\widehat{\sigma}^2}{\pa\rho_d}-\frac{1}{2}\tr\left({\bf C}_{\mathfrak D}^{-1}\frac{\pa{\bf C}_{\mathfrak D}}{\pa\rho_d}\right)-\frac{1}{2}\tr\left({\bf B}_{\mathfrak D}^{-1}\frac{\pa{\bf B}_{\mathfrak D}}{\pa\rho_d}\right)\\
\frac{\pa{\bf B}_{\mathfrak D}}{\pa\rho_d} =& -\tp{\bf H}_{\mathfrak D}{\bf C}_{\mathfrak D}^{-1} \frac{\pa{\bf C}_{\mathfrak D}}{\pa\rho_d} {\bf C}_{\mathfrak D}^{-1}{\bf H}_{\mathfrak D}\\
\frac{\pa{\bf P}_{\mathfrak D}}{\pa\rho_d} =& -{\bf B}_{\mathfrak D}^{-1}\frac{\pa{\bf B}_{\mathfrak D}}{\pa\rho_d}{\bf B}_{\mathfrak D}^{-1}\tp{\bf H}_{\mathfrak D}{\bf C}_{\mathfrak D}^{-1}-{\bf B}_{\mathfrak D}^{-1}\tp{\bf H}_{\mathfrak D}{\bf C}_{\mathfrak D}^{-1}\frac{\pa{\bf C}_{\mathfrak D}}{\pa\rho_d} {\bf C}_{\mathfrak D}^{-1}\\
=& {\bf B}_{\mathfrak D}^{-1} \tp{\bf H}_{\mathfrak D}{\bf C}_{\mathfrak D}^{-1} \frac{\pa{\bf C}_{\mathfrak D}}{\pa\rho_d} {\bf C}_{\mathfrak D}^{-1}{\bf H}_{\mathfrak D} {\bf P}_{\mathfrak D}-{\bf P}_{\mathfrak D}\frac{\pa{\bf C}_{\mathfrak D}}{\pa\rho_d} {\bf C}_{\mathfrak D}^{-1}\\
=& {\bf P}_{\mathfrak D} \frac{\pa{\bf C}_{\mathfrak D}}{\pa\rho_d} {\bf C}_{\mathfrak D}^{-1}[{\bf H}_{\mathfrak D} {\bf P}_{\mathfrak D}-{\bf I}]=-{\bf P}_{\mathfrak D} \frac{\pa{\bf C}_{\mathfrak D}}{\pa\rho_d} {\bf Q}_{\mathfrak D}\\
\frac{\pa{\bf Q}_{\mathfrak D}}{\pa\rho_d} =& -{\bf C}_{\mathfrak D}^{-1} \frac{\pa{\bf C}_{\mathfrak D}}{\pa\rho_d} {\bf C}_{\mathfrak D}^{-1} [{\bf I}-{\bf H}_{\mathfrak D}{\bf P}_{\mathfrak D}]-{\bf C}_{\mathfrak D}^{-1} {\bf H}_{\mathfrak D} \frac{\pa{\bf P}_{\mathfrak D}}{\pa\rho_d}\\
=& -{\bf C}_{\mathfrak D}^{-1} \frac{\pa{\bf C}_{\mathfrak D}}{\pa\rho_d} {\bf Q}_{\mathfrak D} +{\bf C}_{\mathfrak D}^{-1} {\bf H}_{\mathfrak D}{\bf P}_{\mathfrak D} \frac{\pa{\bf C}_{\mathfrak D}}{\pa\rho_d} {\bf Q}_{\mathfrak D} = -{\bf Q}_{\mathfrak D}\frac{\pa{\bf C}_{\mathfrak D}}{\pa\rho_d}{\bf Q}_{\mathfrak D}\\
\frac{\pa\widehat{\sigma}^2}{\pa\rho_d} =& (n-q-2)^{-1}\tp{\bf u}_{\mathfrak D} \frac{\pa{\bf Q}_{\mathfrak D}}{\pa\rho_d} {\bf u}_{\mathfrak D}
=-(n-q-2)^{-1}\tp{\bf u}_{\mathfrak D} {\bf Q}_{\mathfrak D}\frac{\pa{\bf C}_{\mathfrak D}}{\pa\rho_d}{\bf Q}_{\mathfrak D} {\bf u}_{\mathfrak D}\\
\frac{\pa l}{\pa\rho_d} =& \frac{n-q}{2(n-q-2)\widehat{\sigma}^2}\tp{\bf u}_{\mathfrak D}{\bf Q}_{\mathfrak D}\frac{\pa{\bf C}_{\mathfrak D}}{\pa\rho_d}{\bf Q}_{\mathfrak D}{\bf u}_{\mathfrak D}-\frac{1}{2}\tr\left({\bf C}_{\mathfrak D}^{-1}\frac{\pa{\bf C}_{\mathfrak D}}{\pa\rho_d}\right)+\frac{1}{2}\tr\left({\bf P}_{\mathfrak D}\frac{\pa{\bf C}_{\mathfrak D}}{\pa\rho_d} {\bf C}_{\mathfrak D}^{-1}{\bf H}_{\mathfrak D}\right)\\
=& \frac{n-q}{2(n-q-2)\widehat{\sigma}^2}\tp{\bf u}_{\mathfrak D}{\bf Q}_{\mathfrak D}\frac{\pa{\bf C}_{\mathfrak D}}{\pa\rho_d}{\bf Q}_{\mathfrak D}{\bf u}_{\mathfrak D}-\frac{1}{2}\tr\left({\bf C}_{\mathfrak D}^{-1}\frac{\pa{\bf C}_{\mathfrak D}}{\pa\rho_d}\right)+\frac{1}{2}\tr\left({\bf C}_{\mathfrak D}^{-1}{\bf H}_{\mathfrak D} {\bf P}_{\mathfrak D}\frac{\pa{\bf C}_{\mathfrak D}}{\pa\rho_d}\right)\\
=& \frac{n-q}{2(n-q-2)\widehat{\sigma}^2}\tp{\bf u}_{\mathfrak D}{\bf Q}_{\mathfrak D}\frac{\pa{\bf C}_{\mathfrak D}}{\pa\rho_d}{\bf Q}_{\mathfrak D}{\bf u}_{\mathfrak D}-\frac{1}{2}\tr\left({\bf Q}_{\mathfrak D}\frac{\pa{\bf C}_{\mathfrak D}}{\pa\rho_d}\right)
\end{aligned}
\end{equation}

\begin{equation}\label{likrho_hess}
\begin{aligned}
\frac{\pa^2 l}{\pa\rho_d\pa\rho_{d'}} =& \frac{n-q}{2\widehat{\sigma}^4}\frac{\pa\widehat{\sigma}^2}{\pa\rho_d}\frac{\pa\widehat{\sigma}^2}{\pa\rho_{d'}}-\frac{n-q}{2\widehat{\sigma}^2}\frac{\pa^2\widehat{\sigma}^2}{\pa\rho_d\pa\rho_{d'}}+\frac{1}{2}\tr\left({\bf Q}_{\mathfrak D}\frac{\pa{\bf C}_{\mathfrak D}}{\pa\rho_d}{\bf Q}_{\mathfrak D}\frac{\pa{\bf C}_{\mathfrak D}}{\pa\rho_{d'}}\right)-\frac{1}{2}\tr\left({\bf Q}_{\mathfrak D}\frac{\pa^2{\bf C}_{\mathfrak D}}{\pa\rho_d\pa\rho_{d'}}\right)\\
=& \frac{n-q}{2(n-q-2)^2\widehat{\sigma}^4}\tp{\bf u}_{\mathfrak D}{\bf Q}_{\mathfrak D}\frac{\pa{\bf C}_{\mathfrak D}}{\pa\rho_d}{\bf Q}_{\mathfrak D}{\bf u}_{\mathfrak D} \tp{\bf u}_{\mathfrak D}{\bf Q}_{\mathfrak D}\frac{\pa{\bf C}_{\mathfrak D}}{\pa\rho_{d'}}{\bf Q}_{\mathfrak D}{\bf u}_{\mathfrak D}\\
& -\frac{n-q}{2(n-q-2)\widehat{\sigma}^2}\tp{\bf u}_{\mathfrak D}{\bf Q}_{\mathfrak D}\left[\frac{\pa{\bf C}_{\mathfrak D}}{\pa\rho_d}{\bf Q}_{\mathfrak D}\frac{\pa{\bf C}_{\mathfrak D}}{\pa\rho_{d'}} + \frac{\pa{\bf C}_{\mathfrak D}}{\pa\rho_{d'}}{\bf Q}_{\mathfrak D}\frac{\pa{\bf C}_{\mathfrak D}}{\pa\rho_d} - \frac{\pa^2{\bf C}_{\mathfrak D}}{\pa\rho_d\pa\rho_{d'}}\right]{\bf Q}_{\mathfrak D}{\bf u}_{\mathfrak D}\\
& +\frac{1}{2}\tr\left({\bf Q}_{\mathfrak D}\left[\frac{\pa{\bf C}_{\mathfrak D}}{\pa\rho_d}{\bf Q}_{\mathfrak D}\frac{\pa{\bf C}_{\mathfrak D}}{\pa\rho_{d'}}-\frac{\pa^2{\bf C}_{\mathfrak D}}{\pa\rho_d\pa\rho_{d'}}\right]\right)
\end{aligned}
\end{equation}

\section{Proof of proposition \ref{gradeff}}\label{apdx:gradeff}
Since both $\hat U(\vect\theta^*)|\tilde{\bf u}_{\mathfrak D}$ and $\hat U(\vect\theta^*)|{\bf u}_{\mathfrak D}$ are unbiased estimators for $U(\vect\theta^*)$, MSPE's are just covariance $\sigma^2{\bf C}^{**}$ as in \eqref{linmap} \citep{sacks89}.
We can write the correlation matrix conditioned on $\tilde{\bf u}_{\mathfrak D}$, denoted $\widetilde{\bf C}^{**}$, as follows
\begin{equation}
\begin{aligned}
\widetilde{\bf C}^{**} 
&=  {\bf C}_{\mathfrak E} - \begin{bmatrix}{\bf H}_{\mathfrak E}&{\bf C}_{\mathfrak E\widetilde{\mathfrak D}}\end{bmatrix} \begin{bmatrix}{\bf 0} & \tp{\widetilde{\bf H}}_{\mathfrak D}\\ \widetilde{\bf H}_{\mathfrak D} & \widetilde{\bf C}_{\mathfrak D}\end{bmatrix}^{-1} \begin{bmatrix}\tp{\bf H}_{\mathfrak E}\\{\bf C}_{\widetilde{\mathfrak D}\mathfrak E}\end{bmatrix}
=: {\bf C}_{\mathfrak E^0} - \begin{bmatrix} \vect\alpha&\vect\beta\end{bmatrix} \begin{bmatrix}{\bf A} & \tp{\bf B} \\ {\bf B} & {\bf D}\end{bmatrix}^{-1} \begin{bmatrix}\tp{\vect\alpha}\\ \tp{\vect\beta}\end{bmatrix}
\end{aligned}
\end{equation}
where $\mathfrak E=\{\vect\theta^*\}$, $\vect\alpha:=\begin{bmatrix}{\bf H}_{\mathfrak E^0}&{\bf C}_{\mathfrak{E^0D^0}}\end{bmatrix}$, $\vect\beta:={\bf C}_{\mathfrak{E^0D^1}}$,
${\bf A}= \begin{bmatrix}{\bf 0} & \tp{\bf H}_{\mathfrak D^0}\\ {\bf H}_{\mathfrak D^0} & {\bf C}_{\mathfrak D^0\mathfrak D^0}\end{bmatrix}$, ${\bf B}:= \begin{bmatrix} {\bf H}_{\mathfrak D^1} & {\bf C}_{\mathfrak D^1\mathfrak D^0} \end{bmatrix}$, ${\bf D}:={\bf C}_{\mathfrak D^1\mathfrak D^1}$.

Denote \emph{Schur complements} of ${\bf A}$ and {\bf D} as ${\bf S}_{\bf A}:={\bf D}-{\bf B}{\bf A}^{-1}\tp{\bf B}$ and ${\bf S}_{\bf D}:={\bf A}-\tp{\bf B}{\bf D}^{-1}{\bf B}$ respectively.
According to block matrix inversion and Sherman-Morrison-Woodbury formula \citep{harville97}, we have
\begin{equation}
\begin{aligned}
\widetilde{\bf C}^{**} 
&=: {\bf C}_{\mathfrak E^0} - \begin{bmatrix} \vect\alpha&\vect\beta\end{bmatrix} \begin{bmatrix}{\bf S}^{-1}_{\bf D} & -{\bf A}^{-1}\tp{\bf B}{\bf S}^{-1}_{\bf A} \\ -{\bf D}^{-1}{\bf B}{\bf S}^{-1}_{\bf D} & {\bf S}^{-1}_{\bf A}\end{bmatrix} \begin{bmatrix}\tp{\vect\alpha}\\ \tp{\vect\beta}\end{bmatrix}\\
&=: {\bf C}_{\mathfrak E^0} - \begin{bmatrix} \vect\alpha&\vect\beta\end{bmatrix} \begin{bmatrix}{\bf A}^{-1}+{\bf A}^{-1}{\tp{\bf B}\bf S}^{-1}_{\bf A}{\bf B}{\bf A}^{-1} & -{\bf A}^{-1}\tp{\bf B}{\bf S}^{-1}_{\bf A} \\ -{\bf D}^{-1}{\bf B}{\bf S}^{-1}_{\bf D} & {\bf S}^{-1}_{\bf A}\end{bmatrix} \begin{bmatrix}\tp{\vect\alpha}\\ \tp{\vect\beta}\end{bmatrix}\\
&= {\bf C}_{\mathfrak E^0} - \vect\alpha{\bf A}^{-1}\tp{\vect\alpha} - \begin{bmatrix} \vect\alpha&\vect\beta\end{bmatrix} \begin{bmatrix}{\bf A}^{-1}{\tp{\bf B}\bf S}^{-1}_{\bf A}{\bf B}{\bf A}^{-1} & -{\bf A}^{-1}\tp{\bf B}{\bf S}^{-1}_{\bf A} \\ -{\bf S}^{-1}_{\bf A}{\bf B}{\bf A}^{-1} & {\bf S}^{-1}_{\bf A}\end{bmatrix} \begin{bmatrix}\tp{\vect\alpha}\\ \tp{\vect\beta}\end{bmatrix}\\
&= {\bf C}^{**} - (\vect\alpha{\bf A}^{-1}\tp{\bf B}-\vect\beta){\bf S}^{-1}_{\bf A}\tp{(\vect\alpha{\bf A}^{-1}\tp{\bf B}-\vect\beta)}
\leq {\bf C}^{**}
\end{aligned}
\end{equation}
where ${\bf S}^{-1}_{\bf A}>0$ because ${\bf S}_{\bf A}={\bf C}_{\mathfrak D^1\mathfrak D^1}- \begin{bmatrix} {\bf H}_{\mathfrak D^1} & {\bf C}_{\mathfrak D^1\mathfrak D^0} \end{bmatrix} \begin{bmatrix}{\bf 0} & \tp{\bf H}_{\mathfrak D^0}\\ {\bf H}_{\mathfrak D^0} & {\bf C}_{\mathfrak D^0\mathfrak D^0}\end{bmatrix}^{-1} \begin{bmatrix}\tp{\bf H}_{\mathfrak D^1}\\{\bf C}_{\mathfrak{D^0\mathfrak D^1}}\end{bmatrix}=\Cor({\bf u}_{\mathfrak D^1}|{\bf u}_{\mathfrak D^0})>0$.
\qed

\section{Algorithms}
\begin{algorithm}[h]
\caption{Mutual Information for Computer Experiments (MICE)}
\label{Alg:MICE}
\begin{algorithmic}
\STATE Given $U(\vect\theta)$, $\vect\Theta$, $\mathcal{GP}({\bf h}(\cdot),{\bf
C}(\cdot,\cdot;\vect\rho))$, $\nu^2$ and $\nu^2_c$, initial design
$(\mathfrak{De},\tilde{\bf u}_{\mathfrak D})$
\STATE{\bf Step 1.} MLE to obtain estimates $\hat{\vect\rho}$ in ${\bf
C}(\cdot,\cdot;\vect\rho))$
\STATE{\bf Step 2.} Fit GP model to $(\mathfrak{De},\tilde{\bf u}_{\mathfrak D})$
\STATE{\bf Step 3.} Generate $\mathfrak{De}^c$ with repect to $\mathfrak{De}$
and then choose $\vect\Theta_{cand}(\subseteq \mathfrak{De}^c)$
\STATE{\bf Step 4.} Solve $\vect\theta^*=\arg\max_{\vect\theta\in\vect\Theta_{cand}} \V(U(\vect\theta;\hat{\vect\rho},\nu^2)|\tilde{\bf u}_{\mathfrak D})/\V(U(\vect\theta;\hat{\vect\rho},\nu^2_c)|\tilde U(\mathfrak{De}^c\backslash\{\vect\theta\}))$
\STATE{\bf Step 5.} Evaluate $\tilde
U(\vect\theta^*)=(U(\vect\theta^*),dU(\vect\theta^*))$ and set $\mathfrak{De}=\mathfrak{De}\cup\{\vect\theta^*\},\tilde{\bf u}_{\mathfrak D}=\tilde{\bf u}_{\mathfrak D}\cup \tilde U(\vect\theta^*)$
\STATE{\bf Step 6.} If the design $\mathfrak{De}$ has reached the desired size,
then stop; otherwise go to step 1
\STATE{Output:} Design $\mathfrak{De}$ and a GP emulator built on it
\end{algorithmic}
\end{algorithm}

\begin{algorithm}[h]
\caption{Adaptive Gaussian Process emulated Geometric Monte Carlo (adpGPeGMC)}
\label{Alg:adpGPeGMC}
\begin{algorithmic}
\STATE Given current design $(\mathfrak{De},\tilde{\bf u}_{\mathfrak D})$,
initialize GP emulator $\mathcal{GP}({\bf h}(\cdot),{\bf C}(\cdot,\cdot;\hat{\vect\rho}))$
\FOR{$n=1$ to $\textrm{Nsamp}$}
\STATE{\bf 1st kernel:} Sample $\vect\theta$ as the current state by GPeGMC.
\STATE{\bf 2nd kernel:} Fit a mixture of Gaussians $q(\cdot)$ with current
design $(\mathfrak{De},\tilde{\bf u}_{\mathfrak D},\mathrm{gFI}_{\mathfrak D})$.
Propose ${\vect\theta}^{*}\sim q(\cdot)$ and accept it with probability $\alpha = \min\left\{1,\frac{\pi({\vect\theta}^{*})/q({\vect\theta}^{*})}{\pi(\vect\theta)/q(\vect\theta)}\right\}$.
\IF{${\vect\theta}^{*}$ accepted}
\STATE Determine if ${\vect\theta}^{*}$ is a regeneration using
\eqref{retroProb} with ${\vect\theta}_{t}=\vect\theta$ and ${\vect\theta}_{t+1}={\vect\theta}^{*}$.
\IF{Regeneration occurs}
\STATE Adapt current design through MICE algorithm \ref{Alg:MICE} and refine the
associated GP emulator.
\STATE Discard ${\vect\theta}^{*}$, sample ${\vect\theta}^{(\ell+1)}\sim
Q(\cdot)$ as in \eqref{TSQ:Q} using rejection sampling.
\ELSE
\STATE Set ${\vect\theta}^{(n+1)} = {\vect\theta}^{*}$.
\ENDIF
\ELSE
\STATE Set ${\vect\theta}^{(n+1)} = \vect\theta$.
\ENDIF
\ENDFOR
\end{algorithmic}
\end{algorithm}

\end{document}